\documentclass[%
 reprint,
superscriptaddress,
 amsmath,amssymb,
 aps,
pre
]{revtex4-2}

\usepackage{graphicx}
\usepackage{dcolumn}
\usepackage{bm}

\usepackage{chngcntr}
\usepackage{graphicx}
\usepackage[margin=0.5in]{geometry}
\usepackage{overpic}
\usepackage{bm}

\newcommand{\wt}{\widetilde}
\newcommand{\eeqref}[1]{equation~\eqref{#1}}
\newcommand{\apref}[1]{Appendix~\ref{#1}}

\newcommand{\e}{\mathrm{e}}
\usepackage{parskip}
\newcommand{\lb}{\left(}
\newcommand{\rb}{\right)}
\usepackage{overpic}
\usepackage{textcomp}
\usepackage{amssymb}
\usepackage{capt-of}
\usepackage{hyperref}
\usepackage{xfrac}
\usepackage{mathtools}
\usepackage{relsize}
\usepackage{tikz}
\usepackage{pstricks}
\usepackage{pstricks-add}
\usepackage{ulem}
\newcommand{\ls}{\left[}
\newcommand{\rs}{\right]}
\newcommand{\lc}{\left\{}
\newcommand{\rc}{\right\}}

\newcommand{\lp}{\left|}
\newcommand{\rp}{\right|}
\newcommand{\acosh}{\mathrm{acosh}}
\newcommand{\acos}{\mathrm{acos}}
\newcommand{\vphi}{\varphi}
\usepackage{xcolor}
\usepackage{pagecolor}
\definecolor{softdark}{HTML}{32302F}
\definecolor{softdarkfg}{HTML}{EBDBB2}

\makeatletter
\newcommand{\globalcolor}[1]{%
  \color{#1}\global\let\default@color\current@color
}

\newcommand*\@dblLabelI {}
\newcommand*\@dblLabelII {}
\newcommand*\@dblLabelIII {}
\newcommand*\@dblLabelIV {}
\newcommand*\@dblequationAux {}

\def\@dblequationAux #1,#2,%
    {\def\@dblLabelI{\label{#1}}\def\@dblLabelII{\label{#2}}}

\def\@qdequationAux #1,#2,#3,#4,%
{\def\@dblLabelI{\label{#1}}
\def\@dblLabelII{\label{#2}}
\def\@dblLabelIII{\label{#3}}
  \def\@dblLabelIV{\label{#4}}}

\def\@tpequationAux #1,#2,#3,%
{\def\@dblLabelI{\label{#1}}
\def\@dblLabelII{\label{#2}}
\def\@dblLabelIII{\label{#3}}}

\newcommand*{\doubleequation}[3][]{%
    \par\vskip\abovedisplayskip\noindent
    \if\relax\detokenize{#1}\relax
       \let\@dblLabelI\@empty
       \let\@dblLabelII\@empty
    \else 
       \@dblequationAux #1,%
    \fi
    \makebox[0.5\linewidth-1.5em]{%
     \hspace{\stretch2}%
     \makebox[0pt]{$\displaystyle #2$}%
     \hspace{\stretch1}%
    }%
    \makebox[0.5\linewidth-1.5em]{%
     \hspace{\stretch1}%
     \makebox[0pt]{$\displaystyle #3$}%
     \hspace{\stretch2}%
    }%
    \makebox[3em][r]{(%
  \refstepcounter{equation}\theequation\@dblLabelI, 
  \refstepcounter{equation}\theequation\@dblLabelII)}%
  \par\vskip\belowdisplayskip
}

\begin{document}

\preprint{APS/123-QED}
\title{Closed-form Solutions to the Dynamics of
  Confined Biased Lattice Random Walks in
Arbitrary Dimensions}

\author{Seeralan Sarvaharman}
 \email{Email: s.sarvaharman@bristol.ac.uk}
\affiliation{Department of Engineering Mathematics, University of Bristol, BS8 1UB, UK}
\author{Luca Giuggioli}%
 \email{Email: Luca.Giuggioli@bristol.ac.uk}
\affiliation{Department of Engineering Mathematics, University of Bristol, BS8 1UB, UK}
\affiliation{Bristol Centre for Complexity Sciences, University of Bristol, BS8 1UB, UK}

\date{15 November 2020}
\begin{abstract}
  Biased lattice random walks (BLRW) are used to model random motion
  with drift in a variety of empirical situations in engineering and
  natural systems such as phototaxis, chemotaxis or gravitaxis. When
  motion is also affected by the presence of external borders
  resulting from natural barriers or experimental apparatuses,
  modelling biased random movement in confinement becomes
  necessary. To study these scenarios, confined BLRW models have been
  employed but so far only through computational
  techniques due to the lack of an analytic framework. Here, we lay
  the groundwork for such an analytical approach by deriving the
  Green's functions, or propagators, for the confined BLRW in arbitrary
  dimensions and arbitrary boundary conditions. By using these
  propagators we construct explicitly the time dependent first-passage
  probability in one dimension for reflecting and periodic domains,
  while in higher dimensions we are able to find its generating
  function. The latter is used to find the mean first-passage passage
  time for a $d$-dimensional box, $d$-dimensional torus or a
  combination of both.  We show the appearance of surprising
  characteristics such as the presence of saddles in the
  spatio-temporal dynamics of the propagator with reflecting
  boundaries, bimodal features in the first-passage probability in
  periodic domains and the minimisation of the mean first-return time
  for a bias of intermediate strength in rectangular
  domains. Furthermore, we quantify how in a multi-target environment
  with the presence of a bias shorter mean first-passage times can be
  achieved by placing fewer targets close to boundaries in contrast to
  many targets away from them. 
\end{abstract}

\maketitle


\section{Introduction}
Random walk models have been ubiquitously applied across a variety of
disciplines both with continuous space-time variables, i.e. Brownian
walks \cite{schillingpartzschbook2012}, and with discrete variables,
i.e. lattice random walks (LRW) \cite{hughesbook1995}. Due to their
simplicity, LRW have been used as null models to understand the
stochastic dynamics in polymer chains \cite{fisher1966}, record
statistics \cite{godrecheetal2017}, population genetics
\cite{ewensbook2004}, foraging behaviour in animals
\cite{okubolevinbook2001}, diffusion on the surface of stars
\cite{lohmarkrug2009}, energy transfer in molecules
\cite{zumofenblumen1982, pearlstein1982}, and protein transport along
DNA \cite{mayoetal2011,shinkolomeisky2019}, to name just a
few. LRW have also inspired many theoretical approaches to study
coverage times \cite{maierbrockmann2017,
  grassberger2017}, resetting random walks \cite{riascosetal2020} and anomaloous dynamics in disordered systems \cite{thielsokolov2016}.

For many real systems, the use of LRW has provided a convenient way to
extract information about the statistics of an important quantity, the
so-called first-passage probability, or a related one, the so-called
first-return probability. They measure the probability that a random
variable has reached or returned to a given value for the first
time. These quantities represent a work-horse in random search
processes and, more generally, in transport calculations
\cite{rednerbook2001, metzleretalbook2014, benichouvoituriez2014}. In
many empirical scenarios when natural or artificial barriers resulting
from experimental apparatuses affect the dynamics, LRW models need to
be modified to account for the presence of boundaries. The effects on
the first passage statistics become quite significant when the spatial
domain is bounded as exemplified by the mean return time (MRT) and
mean first-passage time (MFPT) becoming finite as compared to infinite
when space is unbounded. Explicit expressions for the MRT have been
established long ago \cite{kac1947a}, those for the MFPT up to 3D for
both rectangular and periodic lattices have been known for some time
\cite{condaminetal2005a, condaminetal2007}, while the analogous ones
in higher dimensional cases have been found more recently
\cite{giuggioli2020}.

Despite the large amount of analytic studies on LRW in confined space
and their related first-passage statistics
\cite{hughesbookvol1_1995,hughesbookvol2_1995,benichouvoituriez2014},
there has been no attempt to generalise the expressions for the MRT or
the MFPT when motion is not completely random but possesses a bias in
some direction, the so-called biased lattice random walks
(BLRW). Similarly there has been no analytic progress for the
first-passage and return probability, with studies on confined BLRW
having been mainly computational \cite{sourjikwingreen2012,
  bergbook1993, jekely2009, sweietal2018}.  This is somewhat
surprising given that there are significant areas of research where
BLRW models have been employed. They include biological systems such
as cell migration due to concentration gradients (chemotaxis)
\cite{sourjikwingreen2012, bergbook1993}, bacteria drifting towards a
light source (phototaxis) \cite{jekely2009} or upwards movement of
single-celled algae in response to gravity (gravitaxis)
\cite{hillhader1997}. In engineering it is worth mentioning the
application of BLRW to study routing protocol for wireless sensor
networks \cite{mabroukietal2007}, to analyse the degradation of
pavement \cite{sweietal2018} and to model field-driven translocation
of tracer particles \cite{valovetal2020}.

With the only closed-form results for BLRW in finite domains
pertaining to the generating function of the 1D propagator with two
absorbing boundaries \cite{godoyfujita1992}, there is a need to
develop a general framework that allows to derive analytically various
transport quantities. Here we are able to do so by extending the LRW
techniques in reference \cite{giuggioli2020} to construct analytically
the confined time-dependent propagator and its generating function for
BLRW in arbitrary dimensions and arbitrary boundary conditions. These
propagators are then used to study first-passage and first-return
statistics and obtain analytic expressions for the MRT and MFPT.

The remainder of the paper is organised as
follows. Section~\ref{sec:time_dep_1d} deals with BLRW in 1D; it
develops a symmetrisation procedure that allows to impose different
boundary conditions and find the propagator generating functions. The
time-dependent propagators are also presented. The derivation of time
dependent first-passage probabilities and mean first-passage times
using the propagator expressions form
Section~\ref{sec:firstpassage_1d}. In Section~\ref{sec:higher_dims} we
treat the problem in higher dimensions, using a hierarchical procedure
to obtain BLRW propagators in arbitrary dimensions and arbitrary
boundary conditions. Using these results we derive the MFPT in
$d$-dimension with reflecting boundaries ($d$-box), periodic
boundaries ($d$-torus) or a mixture of periodic and reflecting
boundaries. Lastly, a summary of the findings are presented in
Section~\ref{sec:conclusion}.

\section{Time Dependent Propagators in One Dimension}
\label{sec:time_dep_1d}
We start by considering the dynamics of a random walker with bias on a
1D infinite lattice. It is conveniently described by utilising two
parameters $q$ and $g$. The parameter $q$ controls the `diffusivity',
with $q = 0$ representing a walker that never moves, while $q = 1$ a
walker that moves at each time step. We take the probability of
jumping to the neighbouring site on the left as
$\frac{q}{2} \lb 1 + g \rb$, while the probability of jumping to the
right as $\frac{q}{2} \lb 1 - g \rb$, and $1 - q$ as the probability
of not moving. The parameter $g$ controls the strength of the
bias. When $g = 0$, the movement is diffusive, whereas the cases
$g = 1$ and $g = -1$ are, respectively, the ballistic limit to the
left and right. The dynamics are governed by the Master equation
\begin{align}
  P(n, t+1) &= (1 - q) P(n, t) + \frac{q}{2}(1 - g) P(n - 1, t) \nonumber
  \\ &+ \frac{q}{2}(1 + g) P(n + 1, t),
\label{eq:p_master_eq}
\end{align}
with $n$ representing the lattice site and $t$ the discrete time
variable. The solution of \eeqref{eq:p_master_eq} can be obtained by
Fourier transforming,
$\widehat{P}(\kappa, t) = \sum^{\infty}_{n = -\infty} P(n, t)
\e^{-\mathrm{i}\kappa n}$, subsequently by finding the generating
function and finally by inverse transforming to real space to obtain
\cite{godoyfujita1992}
\begin{align}
\wt{P}_{n_0}(n , z) = 
\frac{\eta f^{\frac{n - n_0}{2}} \alpha^{-\left|n - n_0 \right|}}
{z q \sinh\ls\acosh\lb {\frac{\eta}{\beta}}\rb\rs},
\label{eq:unbounded_sol}
\end{align}
with $\wt{P}(n, z) = \sum_{t = 0}^{\infty} P(n, t) z^t$ and with $n_0$
indicating the localised initial condition
$P(n, 0) = \delta_{n, n_0}$, where $\delta$ is a Kronecker delta. For
convenience we have employed the following notation:
\begin{align}
&f = \frac{1 - g}{1 + g},
\quad \ \ \ \ \ \ \ \ \ \eta = \frac{1 + f}{2 \sqrt{f}}, \nonumber \\
&\beta = \frac{z q}{1 - z \lb 1 - q \rb},
\quad \alpha = \exp\ls \acosh\lb \frac{\eta}{\beta}\rb \rs.
\end{align}
and the subscript notation $P_{n_0}$ to denote a Kronecker delta
initial condition. The absence of a bias, that is $g \to 0$, implies
that $f, \eta \to 1$, and one recovers the expression of the propagator of the
so-called lazy lattice walker \cite{giuggioli2020}, that is a Polya's
walk where the walker may also stay put at each time step.

\subsection{Symmetrisation Procedure in Presence of Boundaries}
\label{sec:symmetrisation}
When imposing boundary conditions, the method of images is an
intuitive and effective technique to solve the Master
equation. However, when the dynamics are spatially asymmetric, the
method breaks down. If one wishes to employ it, the Master equation
needs to be made symmetric first. This can be accomplished using a
technique used originally by Montroll \cite{montroll1967}. That
technique was used to construct the propagator for a biased
continous-time random walk in presence of a single boundary. Here we
extend that technique to multiple boundaries. Applying the
transformation
\begin{align}
  \label{eq:q_transform}
  Q(n, t) = f^{-\frac{n}{2}}P(n, t)\omega^t - \mu f^{-\frac{n+1}{2}}P(n+1, t) \omega^t
\end{align}
to \eeqref{eq:p_master_eq}, or applying its equivalent in $z$-domain
\begin{align}
\wt{Q}(n, z) = f^{-\frac{n}{2}}\wt{P}(n, z \omega) - \mu f^{-\frac{n + 1}{2}}\wt{P}(n+1, z \omega),
  \label{eq:q_transform_z}
\end{align}
where $\mu \geq 0$ and $\omega^{-1} = 1 - q + \frac{q}{\eta}$, results
in a symmetrised dynamics given by
\begin{align}
  Q(n, t+1) &= \omega \lb 1 - q \rb Q(n, t) \nonumber \\
  &+ \frac{q \omega}{2 \eta}\ls\bigg. Q(n - 1, t) + Q(n + 1, t) \rs.
\label{eq:q_master_eq}
\end{align}

To transform back from the symmetric probability $\wt{Q}(n, z)$
to the original $\wt{P}(n, z)$, one exploits the recursive nature of
transformation \eqref{eq:q_transform_z} to write
\begin{align}
  \label{eq:inverse_q_transform}
  \wt{P}(n, z)  = f^{\frac{n }{2}}
  \sum_{j = 0}^{\infty}
  \mu^j
\wt{Q}\lb m + j, \frac{z}{\omega} \rb,
\end{align}
where $\wt{Q}\lb n, z \rb$ is the general solution to
\eeqref{eq:q_master_eq} in $z$-domain. The corresponding initial
condition of $Q(n,t)$ is related to that of $P(n,t)$ via
$Q(n, 0) = f^{-\frac{n}{2}} P(n, 0) - \mu f^{-\frac{n + 1}{2}} P(n,
0)$. The general solution of \eeqref{eq:q_master_eq} is given by
\begin{align}
  \label{eq:q_sol_z}
  \wt{Q}(n, z) = \sum_{m = -\infty}^{\infty}  Q(m, 0) \wt{H}_{m}(n, z).
\end{align}
where \cite{giuggioli2020}
\begin{align}
  \label{eq:hfunc}
\wt{H}_{n_0}(n, z) = 
\frac{\eta \vphi ^{-\left| n - n_0 \right|} }
{z \omega q \sinh\ls \acosh\lb \frac{1}{\zeta}\rb \rs}.
\end{align}
is the propagator of \eeqref{eq:q_master_eq} and with 
\begin{align}
  \label{eq:zeta}
\zeta = \frac{z \omega q}{\eta \ls 1 - z \omega \lb 1- q \rb
    \rs}
\end{align}
and
\begin{align}
  \label{eq:phi}
\vphi = \exp\ls \acosh \lb \frac{1}{\zeta} \rb \rs.
\end{align}

Using the symmetric solution \eqref{eq:hfunc}, it becomes possible to
apply the method of images for various types of boundary
conditions. In the following sections, to distinguish the different
cases, we use the calligraphic notation, i.e.
$\mathcal{P}^{(\gamma)}$ and $\mathcal{Q}^{(\gamma)}$, for
semi-bounded domains, and $P^{(\gamma)}$ and $Q^{(\gamma)}$ for finite
domains where $\gamma = a, r, m, p$ represents, respectively,
absorbing, reflecting, mixed (one reflecting and one absorbing) and
periodic boundary conditions. The unbounded occupation probability is
represented by $P$ and $Q$ without any superscript $\gamma$.

\subsection{Semi-bounded Propagators}
\begin{figure*}[!ht]
\centering
\begin{overpic}[tics=4, percent,width=0.65\textwidth]{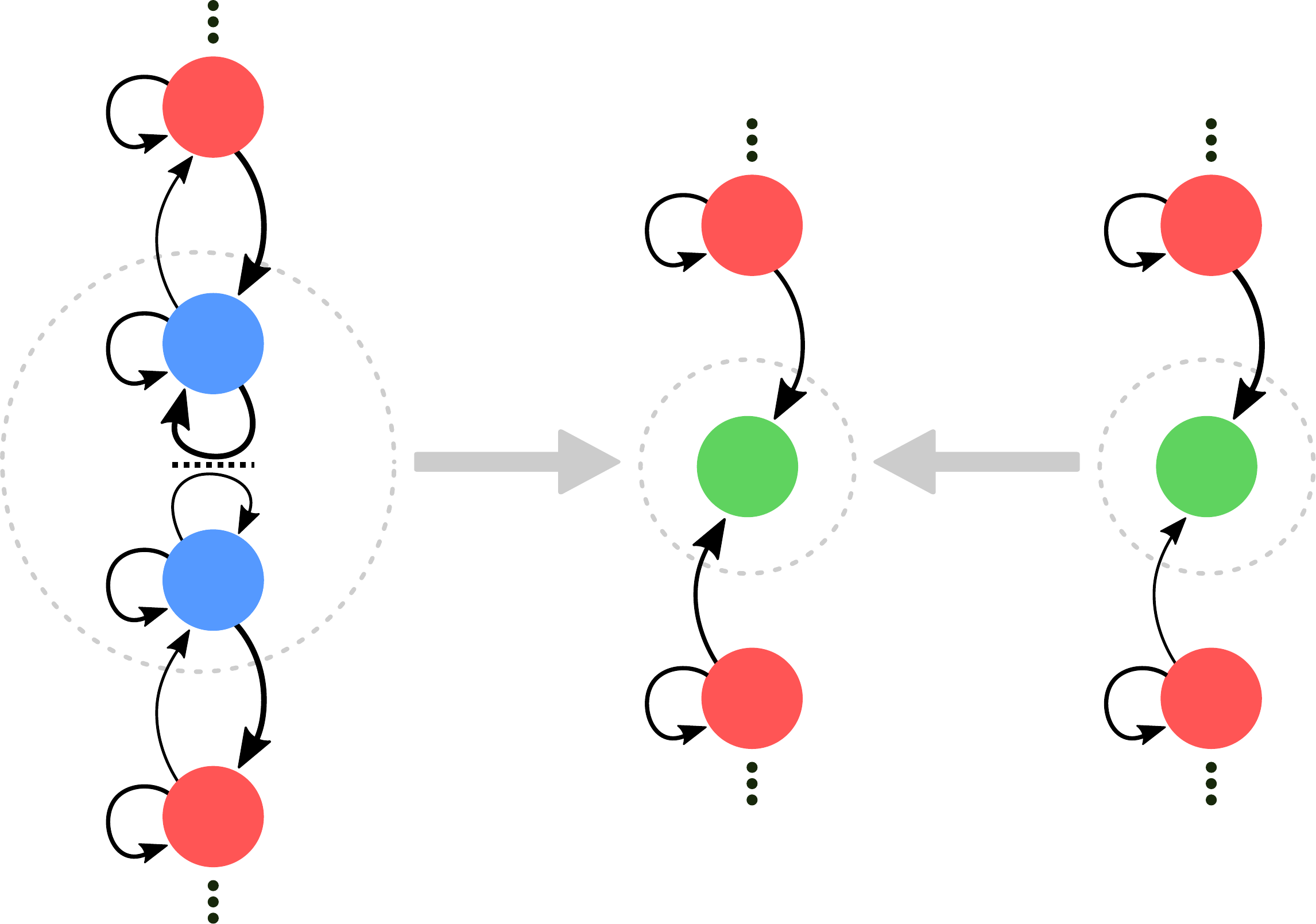}
\put(13.60, 61.5){\color{black}\fontsize{8}{8}$n+2$}
\put(13.60, 7.5){\color{black}\fontsize{8}{8}$n-1$}
\put(15.5, 25.5){\color{black}\fontsize{8}{8}$n$}
\put(13.60, 43.5){\color{black}\fontsize{8}{8}$n+1$}

\put(54.40, 16.5){\color{black}\fontsize{8}{8}$n-1$}
\put(56.10, 34.2){\color{black}\fontsize{8}{8}$n$}
\put(54.40, 52.6){\color{black}\fontsize{8}{8}$n+1$}

\put(89.3, 16.5){\color{black}\fontsize{8}{8}$n-1$}
\put(89.3, 52.4){\color{black}\fontsize{8}{8}$n+1$}
\put(91.0, 34.2){\color{black}\fontsize{8}{8}$n$}

\put(-6, -8){\parbox{0.33\textwidth}{\centering Reflecting\\Boundary
    \\(asymmetric dynamics)}}

\put(35, -8){\parbox{0.33\textwidth}{\centering Absorbing\\Boundary
    \\(symmetrised dynamics)}}

\put(70, -8){\parbox{0.33\textwidth}{\centering Absorbing\\Boundary
    \\(asymmetric dynamics)}}

\put(2, 61){\fontsize{8}{8}$1 - q$}
\put(2, 52){\fontsize{8}{8}$\frac{q}{2} \lb 1 - g \rb$}
\put(21.5, 52){\fontsize{8}{8}$\frac{q}{2} \lb 1 + g \rb$}
\put(1, 43){\colorbox{white} {\fontsize{8}{8}$1 - q$}}
\put(1, 25){\colorbox{white}{\fontsize{8}{8}$1 -q$}}

\put(20, 31){\colorbox{white} {\fontsize{8}{8}$\frac{q}{2} \lb 1 - g \rb$}}
\put(20, 37.5){\colorbox{white} {\fontsize{8}{8}$\frac{q}{2} \lb 1 + g \rb$}}
\put(2, 7){\fontsize{8}{8}$1 - q$}

\put(2, 16){\fontsize{8}{8}$\frac{q}{2} \lb 1 - g \rb$}
\put(21.5, 16){\fontsize{8}{8}$\frac{q}{2} \lb 1 + g \rb$}

\put(39.2, 52.2){\fontsize{8}{8}$\omega \lb 1 - q \rb$}
\put(39.2, 16.2){\fontsize{8}{8}$\omega \lb 1 - q \rb$}

\put(77.7, 52.2){\fontsize{8}{8}$1 - q$}
\put(77.7, 16.2){\fontsize{8}{8}$1 - q$}

\put(77.5, 24){\fontsize{8}{8}$\frac{q}{2} \lb 1 - g \rb$}
\put(97, 44){\fontsize{8}{8}$\frac{q}{2} \lb 1 + g \rb$}

\put(49, 24){\fontsize{8}{8}$\frac{q\omega}{2 \eta} $}
\put(62, 44){\fontsize{8}{8}$\frac{q\omega}{2 \eta}$}
\end{overpic}
\vspace{60pt}
\caption{(Colour Online) Schematic diagram showing the effect of the symmetrising
  transformation \eqref{eq:q_transform} or \eqref{eq:q_transform_z} on
  a single reflecting (leftmost) or a single absorbing (rightmost)
  boundary with the lattices displayed vertically. The red circles
  represent sites in the bulk of the domain, the blue circles
  represent sites adjacent to a reflecting boundary shown as a dashed
  black line and the green circles are absorbing sites. A reflecting
  boundary is a constraint imposed on two sites: with the barrier
  between the sites $n$ and $n+1$, the flux between them must be zero.
  In so doing the probability of not moving at these sites becomes
  $1 - \frac{q}{2}(1 - g)$ or $1 - \frac{q}{2} (1 + g)$ for the site
  above the boundary $n + 1$ or below the boundary $n$,
  respectively. An absorbing boundary is a constraint on a single
  lattice site where the probability at the site must be zero. Under
  the transformation, both the reflective boundary between sites $n$
  and $n+1$ and the absorbing boundary at $n$ with asymmetric dynamics
  become an absorbing site $n$ with symmetric dynamics.}
\label{fig:schematic_boundary}
\end{figure*}
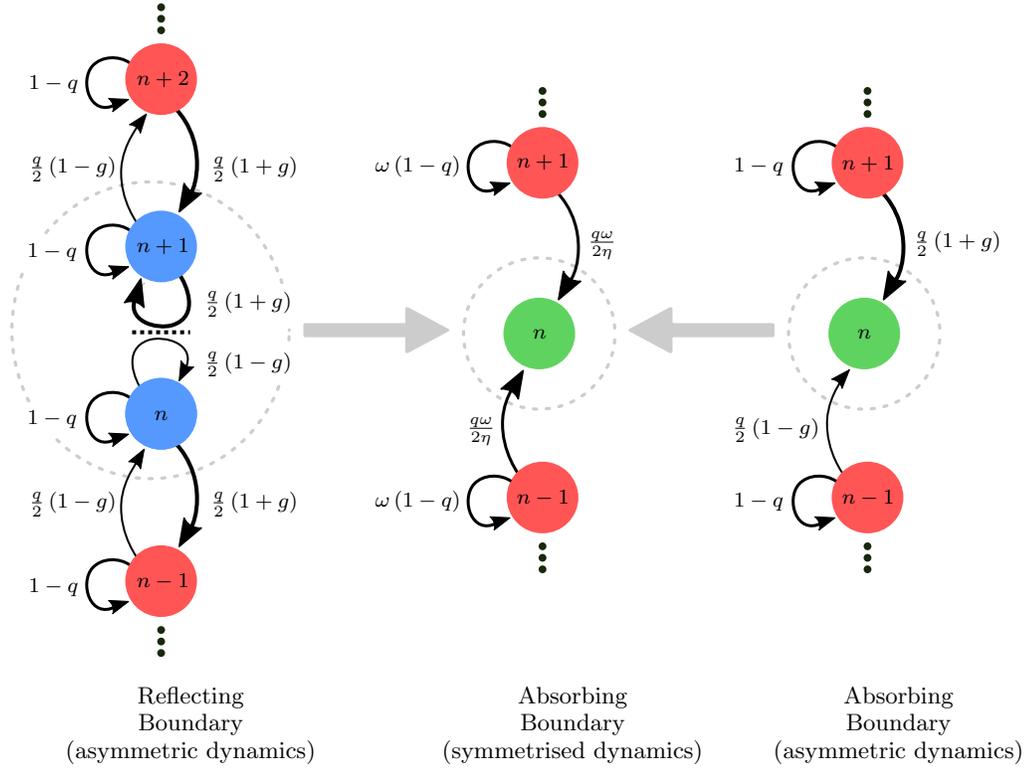

For semi-infinite domains we consider bias random walks on
$\mathbb{Z}^+$. The two straightforward types of boundary conditions
that one can impose are a single reflection and a single absorption;
they are pictorially represented in
figure~\ref{fig:schematic_boundary}. In both of these cases, the
semi-bounded propagator is constructed as a superposition of two
unbounded propagators.  For a single absorbing boundary at $n = 1$,
the requirement $\wt{P}(1, z) = 0$ corresponds, in the symmetric
propagator, to $\wt{Q}(1, z) = 0$ and with $\mu = 0 $. The boundary
condition is satisfied using a single mirror image giving the general
solution
$\wt{\mathcal{Q}}^{(a)}(n, z) = \sum_{m = 1}^{\infty}
\mathcal{Q}^{(a)}(m, 0) \ls\big.  \wt{H}_{m}(n, z) - \wt{H}_{2-m}(n,
z) \rs $, where the spatial convolution is over the semi-infinite
domain and where $\mathcal{Q}^{(a)}(m,0)$ is the initial condition
after symmetrisation, that is obtained from \eeqref{eq:q_transform}
when $t=0$. For an initial condition
$\mathcal{P}^{(a)}(n,0)=\delta_{n,n_0}$ the propagator with a single
absorbing boundary at site $n=1$ is
\begin{align}
\label{eq:p_bounded_a_1}
\wt{\mathcal{P}}_{n_0}^{(a)}(n , z) = 
\frac{\eta f^{\frac{n - n_0}{2}} \lb  \alpha^{-\left|n - n_0 \right|}-\alpha^{-\left|n + n_0  - 2 \right|} \rb}
{z q \sinh\ls\acosh\lb {\frac{\eta}{\beta}}\rb\rs}.
\end{align}
A reflective boundary condition on the asymmetric propagator requires
the flux across the boundary to be zero. With the boundary between
site $n = 0$ and $n = 1$, the zero flux condition is given by
$ f \wt{P}(0 , z) - \wt{P}(1 , z) = 0 $. The corresponding conditions
on the symmetric propagator are $\mu = f^{-\frac{1}{2}}$ and
$\wt{Q}(0, z) = 0$.  With $\mu \neq 0$, under the transformation
\eqref{eq:q_transform_z}, the space between the lattice in the $P$
domain become sites in the $Q$ domain and vice versa. The zero flux
boundary condition is transformed into an absorbing one that is
satisfied using a single image and a symmetrised initial condition
$\mathcal{Q}^{(a)}(n, 0)$, i.e
$\wt{\mathcal{Q}}^{(a)}(n, z) = \sum_{m = 0}^{\infty}
\mathcal{Q}^{(a)}(m, 0)\ls \wt{H}_{m}(n, z) - \wt{H}_{-m}(n, z)
\big.\rs$, where once again $\mathcal{Q}^{(a)}(m, 0)$ is the initial
condition obtained from \eeqref{eq:q_transform} when
$t=0$. Transforming back to the original propagator using
\eeqref{eq:inverse_q_transform} is quite involved and key steps are
given in \apref{sec:symmterisation_single}. For an initial condition
$\mathcal{P}^{(r)}(n, 0) = \delta_{n, n_0}$ the propagator with a
single reflective boundary between sites $n = 0$ and $n = 1$ is given
by
\begin{align}
\label{eq:p_bounded_r_1}
\wt{\mathcal{P}}^{(r)}_{n_0}(n, z) =
\frac{\eta f^{\frac{n - n_0}{2}}
  \lb \alpha^{-\lp n - n_0 \rp}  - \alpha^{-\lp  n + n_0 \rp} \xi \rb}
{z q \sinh\ls\acosh\lb {\frac{\eta}{\beta}}\rb\rs},
\end{align}
where 
\begin{align}
\xi = \frac{f^{\frac{1}{2}} -  \alpha}{f^{\frac{1}{2}}- \alpha^{-1}}
\end{align}

In figure~\ref{fig:schematic_boundary} we display pictorially the two
transformations in the absorbing and reflecting cases, both leading to
an absorbing boundary condition in the symmetrised case.

\subsection{Bounded Propagators}
\label{sec:b_prop}
Having studied propagators on a semi-infinite domain we now turn to
random walks on the finite 1D lattice $1 \leq n \leq N$. We start with
the simplest of these cases, a finite domain with two absorbing
walls. In addition to the absorbing site $n = 1$ we have an absorbing
boundary at $n = N$ giving the further constraint $\wt{P}( N, z) = 0$,
which corresponds to the condition $\wt{Q}( N, z) = 0$ and $\mu = 0$
in \eeqref{eq:q_transform_z}. In this case, the bounded solution is
constructed with infinite images of the unbounded propagators and the
convolution is only over sites within the domain,
\begin{widetext}
$\wt{Q}^{(a)}(n, z) = \sum_{m = 1}^{N} \sum_{k = -\infty}^{+\infty}
Q^{(a)}(m, 0) \ls \big. \wt{H}_{m - 2k\lb N-1\rb}(n, z)  - \wt{H}_{2-m
  - 2k\lb N-1\rb}(n, z) \rs$, where $Q^{(a)}(m,0)$ is the symmetrised
initial condition in a finite domain obtained from
\eeqref{eq:q_transform}.  With a localised initial condition,
$P^{(a)}(n, 0) = \delta_{n, n_0}$, after computing the double
summation the propagator with two absorbing boundaries is
\begin{align}
\label{eq:p_bounded_a_1N}
\wt{P}^{(a)}_{n_0}(n, z) = 
\frac{\eta f^{\frac{n - n_0}{2}}}{z q \sinh\ls \acosh \lb \frac{\eta}{\beta} \rb \rs}
\lc \frac{2 \sinh \ls \lb N - n_> \rb \acosh\lb \frac{\eta}{\beta} \rb \rs  \sinh \ls \lb n_< - 1 \rb \acosh\lb \frac{\eta}{\beta} \rb \rs }
{\sinh\ls \lb N -1  \rb \acosh \lb \frac{\eta}{\beta} \rb \rs} \rc,
\end{align}
where we use the notation $n_> = \frac{1}{2}\lb n + n_0 + \lp n - n_0 \rp\rb$ and
$n_< = \frac{1}{2}\lb n + n_0 - \lp n - n_0 \rp\rb$.

For two reflective boundaries we consider a domain with two
impenetrable barriers: the first between the sites $n = 0$ and
$n = 1$, the second between the sites $n = N$ and $n = N+1$, thus
imposing the constraints $f\wt{P}(0 , z) - \wt{P}( 1 , z) = 0$ and
$f\wt{P}(N, z) - \wt{P}(N+1 , z) = 0$, respectively. With the choice
$\mu = f^{-\frac{1}{2}}$ in \eeqref{eq:q_transform_z}, these
constraints corresponds to the conditions $\wt{Q}( 0, z)$ = 0 and
$\wt{Q}( N, z)$ on the symmetric propagator We follow the same
procedure as the absorbing case by constructing the bounded solution
with infinite images of the unbounded propagator,
$\wt{Q}^{(a)}(n, z) = \sum_{m = 0}^{N} \sum_{k = -\infty}^{+\infty}
Q^{(a)}(m , 0) \ls \big. \wt{H}_{m + 2kN}(n, z) - \wt{H}_{-m + 2kN}(n,
z) \rs$, where once again $Q^{(a)}(m,0)$ is the initial condition
after symmetrisation from \eeqref{eq:q_transform} (for a full
derivation see \apref{sec:symmetrisation_two_bound}).  With the initial
condition $P^{(r)}(n, 0) = \delta_{n, n_0}$, the resulting propagator
with two reflective boundaries is
\begin{align}
\label{eq:p_bounded_r_1N}
\wt{P}_{n_0}^{(r)} (n, z) &= 
  \frac{f^{\frac{n-n_0 - 1}{2}}
\left\{f^{\frac{1}{2}} \sinh{\left[\lb N - n_{>} \rb \acosh\left(\frac{\eta}{\beta} \right) \right]} -
      \sinh{\left[\left(N+1 - n_{>}\right) \acosh\left(\frac{\eta}{\beta} \right) \right]}\right\}}
  {\lb z - 1 \rb \sinh\ls \acosh \left( \frac{\eta}{\beta} \right) \rs
    \sinh\ls N \acosh \left( \frac{\eta}{\beta} \right) \rs} \nonumber \\
&\times \left\{f^{\frac{1}{2}} \sinh{\left[n_{<} \acosh\left(\frac{\eta}{\beta} \right) \right]} -
      \sinh{\left[\left(n_{<} - 1\right) \acosh\left(\frac{\eta}{\beta} \right) \right]}\right\} .
\end{align}

For the mixed boundary condition (reflecting between $n=0$ and $n = 1$
and absorbing at $n = N$) we take the propagator with a single
reflective boundary given in \eeqref{eq:p_bounded_r_1}, and construct
the propagator by considering the probability of being at site $n$ and
having not visited the boundary site $N$ \cite{giuggiolietal2019},
$P_{n_0}^{(m)}\lb n, t \rb = \mathcal{P}_{n_0}^{(r)}(n, t) - \sum_{t'
  = 0}^{t} F^{(r)}_{n_0}(N, t') \mathcal{P}^{(r)}_{N}(n, t-t')$, where
$F^{(r)}_{n_0}(n, t)$ is the first-passage probability of being at
site $n$ at time $t$ for a walker that started at site $n_0$ in a
lattice with an impenetrable barrier between $n=0$ and $n=1$. In
$z$-domain the relation is simply
$\wt{P}_{n_0}^{(m)}(n, z) = \wt{\mathcal{P}}_{n_0}^{(r)}(n, z) -
\wt{F}^{(r)}_{n_0}(N, z) \wt{\mathcal{P}}_{N}^{(r)}(n, z)$ where
$\wt{F}^{(r)}_{n_0}(N, z)$ can be found in \eeqref{eq:fpz_1d_r}. After
some algebra one finds the expression
\begin{align}
  \label{eq:p_bounded_m}
  \wt{P}_{n_0}^{(m)} (n, z) = 
\frac{2 f^{\frac{n-n_0}{2}} \eta \sinh{\left[\left(N - n_{>}\right) \acosh\left(\frac{\eta}{\beta} \right) \right]}
\left\{f^{\frac{1}{2}} \sinh{\left[n_{<} \acosh\left(\frac{\eta}{\beta} \right) \right]} -
\sinh{\left[\left(n_{<} - 1\right) \acosh\left(\frac{\eta}{\beta} \right) \right]}\right\}}
{ z q \sinh\ls \acosh \left( \frac{\eta}{\beta} \right) \rs \left\{f^{\frac{1}{2}} \sinh{\left[N
\operatorname{acosh}{\left(\frac{\eta}{\beta} \right)} \right]}
- \sinh{\left[\left(N - 1\right) \acosh{\left(\frac{\eta}{\beta} \right)} \right]}\right\}}.
\end{align}

Lastly, we consider a biased random walk on a 1D periodic domain with
$N$ distinct sites, which implies that
$\wt{P}(n, z) = \wt{P}(n + kN, z)$ for any integer $k$. To satisfy the
boundary condition one simply wraps the unbounded propagator
\eqref{eq:unbounded_sol} via the summation
$\wt{P}_{n_0}^{(p)}(n, z)=\sum_{k = -\infty}^{\infty} \wt{P}_{n_0}(n +
kN, z)$. Evaluating the sum yields
\begin{align}
\wt{P}_{n_0}^{(p)}(n, z) = \frac{\eta f^{\frac{n - n_0}{2}}
\lc \sinh\ls \lb  N - \lp n - n_0\rp \rb \acosh\lb \frac{\eta}{\beta} \rb \rs
  + f^{-\frac{N \, \mathrm{sgn}(n - n_0)}{2}} \sinh\ls  \lp n - n_0\rp  \acosh{\lb\frac{\eta}{\beta}\rb}\rs \rc}
  {z q \sinh\ls \acosh{\lb \frac{\eta}{\beta}\rb} \rs
    \lb \cosh\ls N \acosh\lb \frac{\eta}{\beta} \rb\rs - \cosh\ls N \acosh{\lb\eta\rb}\rs \rb}, 
\label{eq:p_bounded_p}
\end{align}
where $\mathrm{sgn}(n)$ is the signum function, defined as
$\mathrm{sgn}(n) = -1$ when $n < 0$, $\mathrm{sgn}(n) = 1$ when
$n > 0$, and $\mathrm{sgn}(n) = 0$ when $n = 0$.

Equations \eqref{eq:p_bounded_r_1N}, \eqref{eq:p_bounded_m} and
\eqref{eq:p_bounded_p} are not known in the literature, even though
expressions similar to equations \eqref{eq:p_bounded_r_1N} and
\eqref{eq:p_bounded_m} can be found in reference
\cite{khanthabalakrishnan1984, khanthabalakrishnan1985a}, where the
continuous time BLRW was derived using an alternative procedure. This
procedure was also used in reference \cite{godoyfujita1992} to derive
equations \eqref{eq:p_bounded_a_1}, \eqref{eq:p_bounded_r_1}, and
\eqref{eq:p_bounded_a_1N}, for the discrete time BLRW, but only for
the case when $q=1$, that is an always moving walker.
\end{widetext}

\subsection{Time Dependent Propagators with Finite Domains}
In order to find the time dependence of the propagators one must
evaluate the integral (inverse $z$ transform)
$P^{(\gamma)}_{n_0}(n, t) = \lb 2 \pi \mathrm{i}\rb^{-1} \oint
\wt{P}^{(\gamma)}_{n_0}(n, z) z^{-t - 1} \mathrm{d}z$, with
$\lp z \rp < 1$ and where the integration contour is
counterclockwise. Equivalently, one can find time dependent solution
more directly by solving the matricial Master equation
\begin{align}
  \label{eq:matrix_master}
\vec{P}(t+1) = \mathbf{A} \cdot \vec{P}( t),
\end{align}
where
\begin{align}
  \mathbf{A} = 
\begin{small}
\begin{bmatrix}
1 - q + \varepsilon & \frac{q}{2}(1 + g)  &        &  &                    & \sigma             \\
\frac{p}{2} (1 - g) & 1 - q               & \ddots &  &                    &                    \\
                    & \frac{q}{2} (1 - g) & \ddots &  & \frac{q}{2}(1 + g) &                    \\
                    &                     & \ddots &  & 1 - q              & \frac{q}{2}(1 + g) \\
\nu                 &                     &        &  & \frac{q}{2}(1 - g) & 1 - q + \delta    
\end{bmatrix}
\end{small}.
\end{align}
\begin{figure}[ht]
  \centering
  \includegraphics[width=0.5\textwidth]{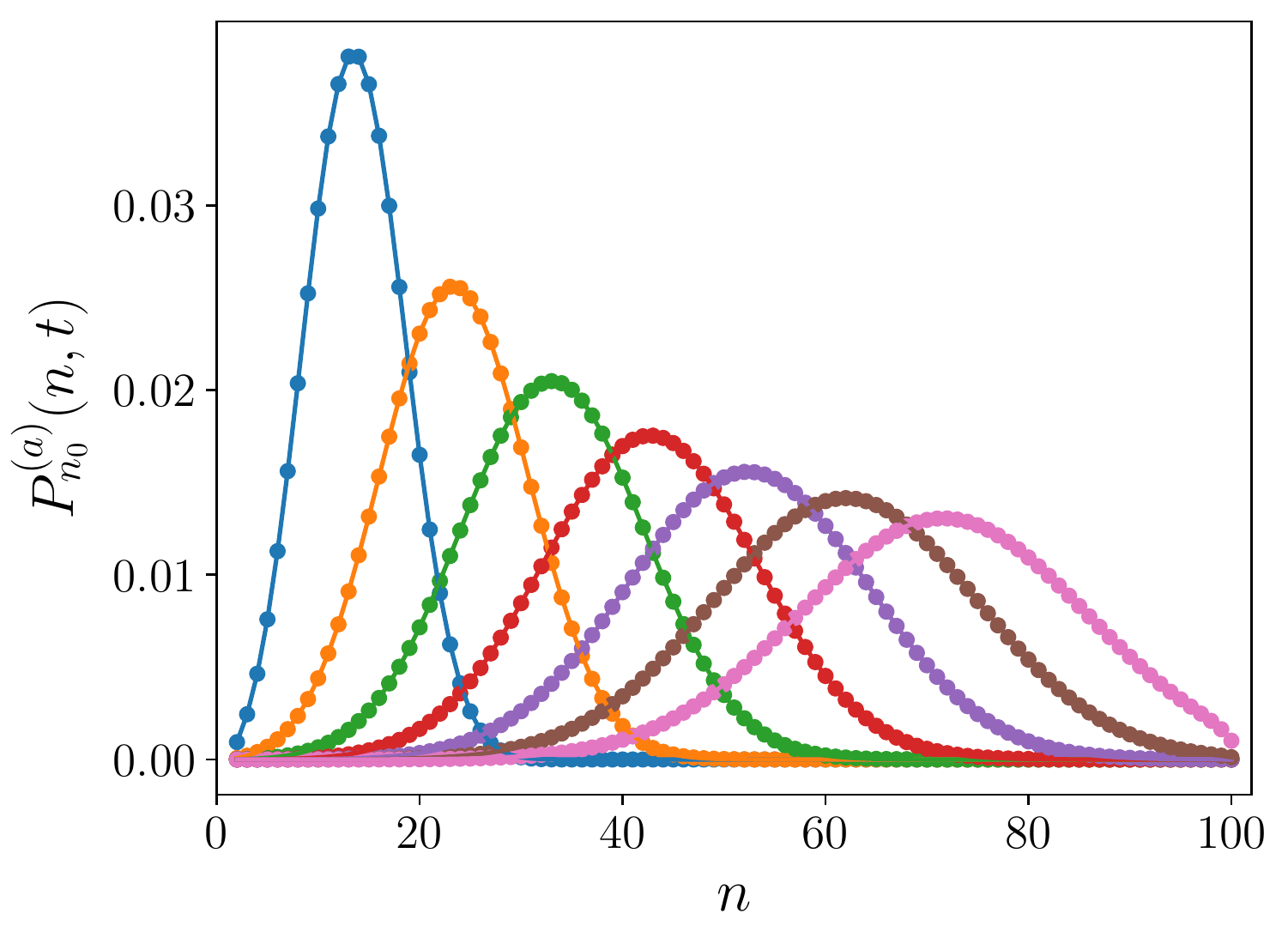}
  \caption{(Colour Online) One dimensional propagator with absorbing boundaries at
    sites $1$ and $101$. The localised initial condition is at
    $n_0 = 2$ and the bias and diffusive parameters are, respectively,
    $g = -0.3$ and $q = 0.8$. Each of the curves represent the
    probability at different times, with the left most curve being at
    $t = 40$, the right most being at $t = 280$ and $\Delta t = 40$
    between each of them. The dots are from \eeqref{eq:1d_all_prop} with $\gamma=a$,
    whereas the solid lines are obtained by solving iteratively
    \eeqref{eq:matrix_master}.}
\label{fig:absorbing_1d_propagator}
\end{figure}
The different types of boundary conditions are accounted for by a
relevant size of $\mathbf{A}$ and appropriately chosen parameters
$\varepsilon, \delta, \sigma$ and $\nu$: reflective boundaries with
$\varepsilon = \frac{q}{2}\lb 1 + g \rb$,
$\delta = \frac{q}{2}\lb 1 - g\rb$, $\nu = \sigma = 0$ and
$\mathbf{A}_{N\times N}$; absorbing boundaries with
$\varepsilon = \delta = \nu = \sigma = 0$ and
$\mathbf{A}_{\lb N-2 \rb \times\lb N-2 \rb}$; mixed boundaries with
$\varepsilon = \frac{q}{2} \lb 1 + g \rb$, $\delta = \nu = \sigma = 0$
and $\mathbf{A}_{\lb N - 1\rb\times \lb N-1\rb}$; and periodic
boundaries with $\nu = \frac{q}{2} \lb 1 + g \rb$,
$\sigma = \frac{q}{2} \lb 1 - g \rb$, $\varepsilon = \delta = 0$ and
$\mathbf{A}_{N\times N}$. By diagonalising the matrix $\mathbf{A}$,
the solution can be written as
$\vec{P}(t) = \mathbf{L} \mathbf{E}^{t} \mathbf{R} \vec{P}( 0)$ where
respectively, $\mathbf{L}$ and $\mathbf{R}$ are matrices containing
the left and right normalised eigenvectors, while $\mathbf{E}$ is the
diagonal matrix of eigenvalues. The spatial dependence is determined
by the eigenvectors while the eigenvalues give the time
dependence. These eigenvalues and eigenvectors are known explicitly
for the absorbing, reflecting and periodic cases~\cite{yueh2005,
  yuehcheng2008, willms2008}, while for the mixed boundary condition
we exploit the properties of Chebyshev polynomials to write a
propagator with time dependent coefficients known numerically (see
\apref{sec:time_dep_mixed} for details). To represent the time dependent
propagator a convenient notation is
\begin{align}
  \label{eq:1d_all_prop}
  P_{n_0}^{(\gamma)}(n, t) = 
  \sum_{k = w^{(\gamma)}}^{W^{(\gamma)}} h_{k}^{(\gamma)}\lb n, n_0 \rb \left[1 + s_k^{(\gamma)}\right]^{t}, 
\end{align}
where $w^{(p)} = 0$ and $W^{(p)} = N-1$ for the periodic case;
$w^{(r)} = 0$ and $W^{(r)} = N-1$ for the reflecting case;
$w^{(a)} = 1 $ and $W^{(a)} = N-2$ for the absorbing case; and
$w^{(m)} = 1 $ and $W^{(m)} = N-1$ for the mixed case. The time
dependence is defined by $\ls 1 + s_k^{(\gamma)} \rs^t$ with
\begin{align}
s_k^{(\gamma)} = \left\{
\begin{array}{cr}
q\cos{\lb \frac{2k\pi}{N}\rb} + \mathrm{i}qg\sin{\lb \frac{2k\pi}{N}\rb} - q, &\gamma=p, \\ [\bigskipamount]
\frac{q}{\eta}\cos{\lb \frac{k \pi}{N - 1} \rb} - q, &\gamma=a, \\ [\bigskipamount]
\frac{q}{\eta}\cos{\lb \theta_k \rb} - q, &\gamma=m, \\ [\bigskipamount]
\left.\begin{array}{cr}
\frac{q}{\eta}\cos{\lb \frac{k \pi}{N} \rb} - q, &k \neq 0, \\ [\bigskipamount]
0, &k = 0,
\end{array}\right\} &\gamma = r,
\end{array}\right.
\label{eq:sk}
\end{align}
where $\cos{\lb \theta_k \rb}$ is the $k^{\mathrm{th}}$ root of the orthogonal
polynomial
$f^{\frac{1}{2}}U_{N-1}\ls \cos{\lb \theta \rb}\rs -
U_{N-2}\ls\cos{\lb \theta \rb}\rs$ and where $U_{n}$ is an
$n^{\text{th}}$ order Chebyshev polynomial of the second kind. The
spatial dependence in \eeqref{eq:1d_all_prop} is
\begin{widetext}
\begin{align}
h_k^{(\gamma)}(n,n_0)=\left\{
\begin{array}{cr}
\frac{\exp\ls \frac{2 k \pi \mathrm{i}(n - n_0)}{N} \rs }{N} , &{\gamma=p}, \\ [\bigskipamount]
  \frac{2f^{\frac{n - n_0}{2}}\sin\ls \lb\frac{n-1}{N-1}\rb k\pi\rs \sin\ls
  \left(\frac{n_0-1}{N-1}\right)k\pi \rs}{N-1}, &{\gamma=a},  \\ [\bigskipamount]
  \frac{2f^{\frac{n - n_0}{2}}\sin \ls \lb N - n_> \rb \theta_k \rs \lc  f^{\frac{1}{2}} \sin \ls  n_<  \theta_k \rs -
  \sin \ls \lb n_<  - 1\rb \theta_k \rs \rc} {
    \lb N - 1 \rb \cos\ls \lb N-1\rb \theta_k\rs - N f^{\frac{1}{2}} \cos{\ls N \theta_k \rs}}, &{\gamma=m}, \\ [\bigskipamount]
  \left.
  \begin{array}{cr}
  \frac{ f^{\frac{n - n_0 - 1}{2}}\lc f^{\frac{1}{2}} \sin \ls \frac{n k \pi}{N} \rs - \sin \ls \lb n - 1\rb
  \frac{k \pi }{N}\rs \rc \lc f^{\frac{1}{2}} \sin \ls \frac{n_0 k \pi}{N} \rs - \sin \ls \lb n_0 - 1\rb
  \frac{k \pi }{N}\rs \rc}{N \lb \eta  - \cos\ls \frac{k \pi}{N} \rs \rb}, &k \neq 0, \\ [\bigskipamount]
  \frac{f^{n - 1 }\lb 1 - f\rb}{1 - f^N}, &k = 0, 
\end{array}\right\} &{\gamma = r}.
\end{array}\right. 
\label{eq:gk}
\end{align}
\end{widetext}
In the periodic case, $h_k^{(p)}(n, n_0)$ and $s_k^{(p)}$ are both
complex, but \eeqref{eq:1d_all_prop} is real. When $g=0$, $s^{(p)}_k$
becomes real and
$h_k^{(p)}(n, n_0) = \cos{\ls \frac{2 k \pi (n - n_0)}{N} \rs}$
because the $\sin$ terms cancels out.

In figure~\ref{fig:absorbing_1d_propagator}, we plot the propagator in
\eeqref{eq:1d_all_prop} with two absorbing boundaries ($\gamma = a$)
at $n = 1$ and at $n = N$ and with a negative bias, $g = -0.3$. The
drift to the right is evident from the movement of the peak of the
probability, while the broadening of the overall shape is due to
diffusion.

Using the 1D propagators in \eeqref{eq:1d_all_prop} it is possible to
recover known solutions to the bounded drift-diffusion equation. In \apref{sec:continuous_limit_ap} we otuline the limiting procedure to obtain the space-time continuous propagators for the four boundary conditions studied.

\begin{widetext}
\section{First-Passage Processes in One Dimension}
\label{sec:firstpassage_1d}
An important quantity in transport calculations, already introduced in
Section \ref{sec:b_prop}, is the first-passage probability,
$F_{n_0}(n,t)$, to reach a target site $n$ from site $n_0$ at time $t$
\cite{montrollweiss1965}. It is directly related to the propagator
through the renewal relation in $z$-domain
$\wt{F}_{n_0}(n, z) = \wt{P}_{n_0}(n, z) / \wt{P}_{n}(n, z)$. We
consider first the reflective domains and subsequently the periodic
domain. Using the propagator \eqref{eq:p_bounded_r_1N}, the generating
function of the first-passage probability is written in a compact
manner (see \eeqref{eq:fpz_1d_r}) by considering the case when
$n > n_0$ and vice versa. Through a $z$-inversion the time dependent
first-passage probability can be written as
\begin{align}
  F^{(r)}_{n_0}(n, t) = \frac{qf^{\frac{n - n_0}{2}}}{\eta} \left\{
\begin{array}{cr}
  \mathlarger{\sum}_{k = 1}^{n-1} \frac{\sin{\lb \theta_k \rb } \lc f^{\frac{1}{2}} \sin{\lb n_0 \theta_k \rb}
  - \sin{\ls \lb n_0 - 1 \rb\theta_k \rs} \rc \ls 1 - p + \frac{p}{\eta} \cos{\lb \theta_k \rb} \rs^{t-1}}
  {\lb n - 1 \rb \cos \ls \lb n - 1\rb \theta_k \rs-f^{\frac{1}{2}} n \cos{\lb n \theta_k \rb} } & n > n_0 \\ [\bigskipamount]
   \mathlarger{\sum}_{k = 1}^{N -n} \frac{\sin{\lb \psi_k \rb} \left( f^{\frac{1}{2}} \sin{\ls \lb N  - n_0 \rb \psi_k \rs}
  - \sin{\ls \lb N+1 - n_0 \rb\psi_k \rs} \right) \ls 1 - p + \frac{p}{\eta} \cos{\lb \psi_k \rb} \rs^{t-1}}
  {\lb N + 1 - n \rb \cos \ls \lb N + 1 -  n \rb \psi_k \rs
  -f^{\frac{1}{2}}\lb N - n \rb  \cos{\ls \lb N - n \rb \psi_k \rs}} & n < n_0,
\end{array} \right.
\label{eq:fpt_1d}
\end{align}
with $F_{n_0}^{(r)}(n, 0) = 0$, and where $\cos\lb \theta_k \rb$ and
$\cos\lb \psi_k \rb$ are, respectively, the $k^{\mathrm{th}}$ roots of
the orthogonal polynomial
$f^{\frac{1}{2}} U_{n-1}\ls \cos\lb \theta_k \rb \rs - U_{n -
  2}\ls\cos\lb \theta_k \rb\rs$ and
$f^{\frac{1}{2}} U_{N-1 -n}\ls \cos\lb \psi_k \rb \rs - U_{N -
  n}\ls\cos\lb \psi_k \rb\rs$ with $U_{-1} = 0$

For the periodic case a similar procedure gives a compact expression
in \eeqref{eq:fpz_1d_p} by treating $n > n_0$ and $n < n_0$
separately. Using the signum function the time dependence can be
written conveniently as the following single expression:
\begin{align}
  \label{eq:fpt_p}
  F_{n_0}^{(p)}(n, t) &= 
    \frac{ q f^{\frac{ n - n_0 }{2}}}{\eta N}\sum_{k = 1}^{N-1}(-1)^{k+1} \sin\lb \frac{k \pi}{N} \rb 
    \lc \sin\ls \lb N - \lp n - n_0 \rp \rb \frac{k \pi }{N}\rs \right. \nonumber  \\
  &+ \left. \sin\ls \lp n - n_0 \rp \frac{k \pi }{N} \rs  f^{-\frac{N \mathrm{sgn}\lb n - n_0\rb}{2}} \rc 
 \ls 1 - q + \frac{q}{\eta}\cos\lb \frac{k\pi}{N} \rb \rs^{t - 1},
\end{align}
with $F_{n_0}^{(p)}(n, 0) = 0$.
\begin{figure*}[!ht]
  \hfill
  \begin{overpic}[tics=4, percent,width=0.95\textwidth]{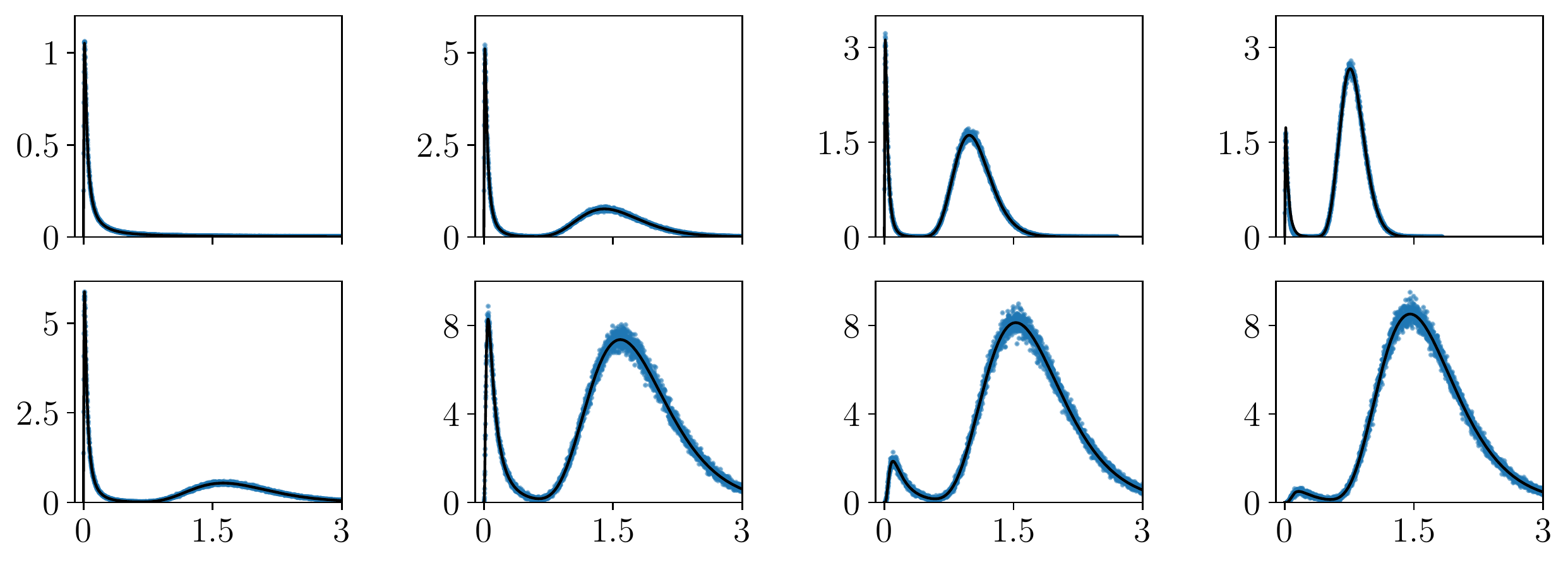}
    \put(-4, 22){\rotatebox{90}{$F^{(p)}_{n_0}(n, t)$}}
    \put(-4, 5){\rotatebox{90}{$F^{(p)}_{n_0}(n, t)$}}

    \put(13, -1.2){$t$}
    \put(17.5, -1.2){$\times 10^3$}
    \put(38.5, -1.2){$t$}
    \put(43, -1.2){$\times 10^3$}
    \put(64, -1.2){$t$}
    \put(68.55, -1.2){$\times 10^3$}
    \put(89.5, -1.2){$t$}
    \put(94, -1.2){$\times 10^3$}

    \put(-2.75, 34){$\times 10^{-2}$}
    \put(-2.75, 17){$\times 10^{-3}$}
    \put(48.5, 34){$\times 10^{-3}$}
    \put(48.5, 17){$\times 10^{-4}$}
    \put(22.75, 34){$\times 10^{-3}$}
    \put(22.75, 17){$\times 10^{-4}$}
    \put(74, 34){$\times 10^{-3}$}
    \put(74, 17){$\times 10^{-4}$}

    \put(17.7, 32){(a)}
    \put(42.9, 32){(b)}
    \put(68.8, 32){(c)}
    \put(94.0, 32){(d)}

    \put(17.7, 15.2){(e)}
    \put(42.9, 15.2){(f)}
    \put(68.8, 15.2){(g)}
    \put(94.0, 15.2){(h)}

  \end{overpic}
  \vspace{10pt}
  \caption{(Colour Online) First-passage probability of a 1D random walk with bias in
    a periodic domain with $N = 50$ sites, initial position
    $n_0 = 2$, and diffusive parameter $q = 0.1$. Panels (a)-(d) all
    have the target at site $n = 4$, but differ in the bias which
    is, respectively, $g = 0, 0.3, 0.45$ and $0.6$. The panels
    (e)-(f) have the same bias, $g = 0.35$, while differing in the
    position of the target site, which is $n = 4, 6, 8$ and $10$,
    respectively. The solid black line is from \eeqref{eq:fpt_p},
    whereas the blue circles are from $10^6$ stochastic simulations.}
\label{fig:fpp_p}
\end{figure*}
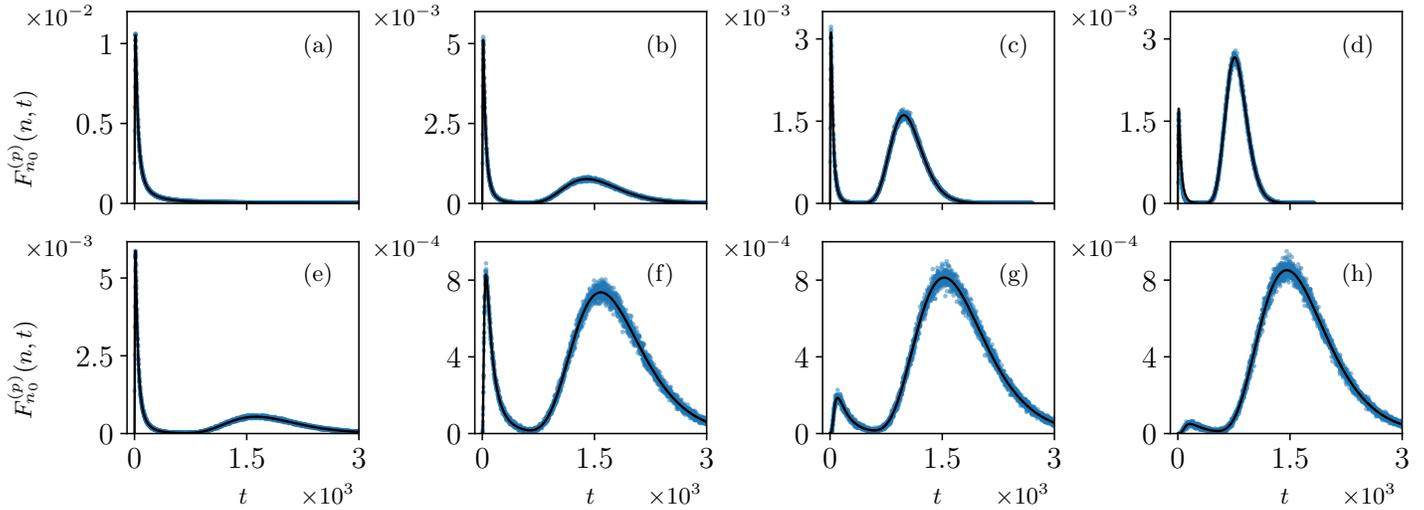
\end{widetext}

In the case of periodic domains, an interesting feature is the
appearance of two peaks in the first-passage probability. While the
first-passage dynamics of a diffusive walker in a periodic domain is
monomodal, in the presence of a bias one can find bimodal features. To
display these features we plot $F_{n_0}^{(p)}(n, t)$ in
figure~\ref{fig:fpp_p}. The panels (a)-(d) depict the first-passage
probability to the same target at $n = 4$ starting from $n_0 = 2$ but
each panel represents a stronger bias from left to right.  The panels
(e)-(h) have the same bias, $g = 0.35$, but the target locations are
displaced away from the starting site $n_0 = 2$. In the absence of a
bias, i.e. figure~\ref{fig:fpp_p}a, one finds a monomodal probability
function characteristic of diffusive processes. As the bias is
increased to positive values (first row), the walker is more likely to
travel leftwards taking the longer route to reach the target (via the
site $N$) resulting in the appearance of a second peak and at the same
time the gradual loss of the first. In the left ballistic limit, one
expects the first peak to be completely lost and the second peak to be
a Kronecker delta at $N - n + n_0$ (see equations
\eqref{eq:fpp_p_nb_m} and \eqref{eq:fpp_p_nb_M} in
\apref{sec:first_passage_ap} for the limiting expressions of
\eeqref{eq:fpt_p} when $f \to 0$ and $f \to \infty$). With the bias
fixed (second row), as one moves the target site further away from
$n_0$ and opposite to the direction of the bias, the first peak is
gradually lost while the second becomes more prominent. With the
target site close to $n_0$, the distance to travel against the bias to
reach $n$ is small enough such that there is still a high probability
of reaching $n$ from $n_0$ without visiting $N$. As one increases the
distance between $n_0$ and $n$, with $n > n_0$, the likelihood of a
walker travelling this distance against the bias decreases resulting
in the progressive loss of the first peak.

\subsection{Mean First-Passage Time}
By using the first-passage generating function with reflecting or
periodic boundaries, in \eeqref{eq:fpz_1d_r} or \eqref{eq:fpz_1d_p} in
\apref{sec:first_passage_ap}, one finds the mean of the probability
through
$T^{(\gamma)}_{n_0 \to n} = \frac{\mathrm{d}}{\mathrm{d}z}
\wt{F}_{n_0}^{(\gamma)}(n, z) {\big|}_{z \to 1}$, to give the MFPT,
\begin{align}
  T^{(r)}_{n_0 \to n} &=\frac{1}{q} \frac{(f + 1)}{(f - 1)^2}
  \lc \big. \lb  n - n_0\rb (f - 1) \vphantom{f^{\frac{N}{2} \ls 1 - \mathrm{sgn} (n - n_0) \rs}} \right.\nonumber \\ 
  &\left. + f^{\frac{N}{2} \ls 1 - \mathrm{sgn} (n - n_0) \rs} \lb f^{1 - n} - f^{1 - n_0}  \rb    \rc
\label{eq:mean_fpt_1d_r}
\end{align}

and
\begin{align}
  T^{(p)}_{n_0 \to n}&=\frac{1}{q} \frac{(f + 1)}{(f - 1) \lb f^N - 1 \rb}
  \lc \lb  n - n_0\rb \lb f^N - 1 \rb \vphantom{N f^{\frac{N}{2} \ls 1 - \mathrm{sgn} (n - n_0) \rs}} \right. \nonumber \\
&\left. + N f^{\frac{N}{2} \ls 1 - \mathrm{sgn} (n - n_0) \rs} \lb 1 - f^{n - n_0}  \rb    \rc
\label{eq:mean_fpt_1d_p}
\end{align}

In the left (right) ballistic limit, that is $q \to 1$ and $f \to 0$
($f \to \infty$) of \eeqref{eq:mean_fpt_1d_r}, we find the MFPT to be
$\lp n - n_0 \rp$ if the target is in the direction of the bias or
infinite if the target is against the bias. On the other hand, the
MFPT with periodic boundaries in \eeqref{eq:mean_fpt_1d_p} will always
be finite: with $n > n_0$ in the left ballistic limit
$T^{(p)}_{n_0 \to n} = N - \lp n - n_0 \rp$, while in the right
ballistic limit $T^{(p)}_{n_0 \to n} = \lp n - n_0 \rp$ and vice
versa. In the diffusive limit, i.e. $f \to 1$, equations
\eqref{eq:mean_fpt_1d_r} and \eqref{eq:mean_fpt_1d_p} reduce to
\cite{giuggioli2020},
$$
T^{(r)}_{n_0 \to n} = \frac{1}{q} \ls \bigg.  N \lp n - n_0 \rp + (n - n_0) (n +
  n_0 - 1 - N)\rs
$$
and
$$
T_{n_0 \to n}^{(p)} = \frac{1}{q} \lb \bigg. N - \lp n - n_0 \rp \rb \lp n - n_0\rp,
$$ 
respectively.

\section{Dynamics in Higher Dimensions}
\label{sec:higher_dims}
To find propagators in higher dimensions we need both the series
solution and the compact solution from the method of images. The
procedure for finding propagators in higher dimension is a slight
variation of the eight step method introduced by one of the present
authors \cite{giuggioli2020}. To illustrate this new procedure we
first present the case of a walker in a 2D domain with reflective boundary
conditions.

\subsection{Two dimensional propagator with reflective
  boundaries}
We start by considering the dynamics of a walker on a 2D lattice that
is bounded along the first dimension whilst unbounded in the
second. The probability of stepping left or right along the first
dimension are, respectively, $\frac{q_1}{4} \lb 1 - g_1 \rb$ or
$\frac{q_1}{4} \lb 1 + g_1 \rb$. Similarly stepping left or right
along the second dimension are, respectively,
$\frac{q_2}{4} \lb 1 - g_2 \rb$ or $\frac{q_2}{4} \lb 1 + g_2 \rb$. In
the bulk of the domain, the probability of remaining at a site is
$1 - \frac{q_1}{2} - \frac{q_2}{2}$, while along the left (or right)
boundary at $n_{_1}= 1$ (or at $n_{_1}= N_1$), is
$1 - \frac{q_1}{4} \lb 1 - g_1 \rb - \frac{q_2}{2}$ (or
$1 - \frac{q_1}{4} \lb 1 + g_1 \rb - \frac{q_2}{2}$). The dynamics in
the bulk of the domain are governed by the Master equation
\begin{align}
\smaller
&P(n_{_1}, n_{_2}, t\! + \!1) = \ls \bigg. 1 - \frac{q_1}{2} - \frac{q_2}{2} \rs P(n_{_1}, n_{_2}, t) \nonumber \\
&+\!\frac{q_1}{4}\ls \bigg. \lb 1\! -\! g_1 \rb P(n_{_1}\!-\! 1, n_{_2}, t) + \lb 1\! +\! g_1 \rb P(n_{_1}\!+\! 1, n_{_2}, t) \rs \nonumber \\
&+\! \frac{q_2}{4}\ls \bigg. \lb 1\! -\! g_2 \rb P(n_{_1}, n_2\! -\! 1, t) + \lb 1\! +\! g_2 \rb P(n_{_1}, n_2\! +\! 1, t) \rs,
\label{eq:master_eq_2d}
\end{align}
along the left boundary by
\begin{align}
&P(1, n_{_2}, t\! +\! 1) = \ls\bigg. 1 - \frac{q_1}{4} \lb 1\! -\! g_1 \rb - \frac{q_2}{2} \rs P(1, n_{_2}, t) \nonumber \\
&+ \frac{q_2}{4}\ls \bigg. \lb 1\! -\! g_2 \rb P(1, n_2 - 1, t) + \lb 1\! +\! g_2 \rb P(1, n_2 + 1, t) \rs \nonumber \\
&+\frac{q_1}{4}\lb 1 + g_1 \rb P(2, n_{_2}, t),
\label{eq:master_eq_2d_lb}
\end{align}
and along the right boundary by
\begin{align}
&P(N_1, n_{_2}, t\! +\! 1) = \ls\bigg. 1 - \frac{q_1}{4} \lb 1\! +\! g_1 \rb - \frac{q_2}{2} \rs P(N_1, n_{_2}, t) \nonumber \\
&+ \frac{q_2}{4}\ls \bigg. \lb 1\! -\! g_2 \rb P(1, n_2 - 1, t) + \lb 1\! +\! g_2 \rb P(1, n_2 + 1, t) \rs \nonumber \\
&+\frac{q_1}{4}\lb 1 - g_1 \rb P(N_1 - 1, n_{_2}, t),
\label{eq:master_eq_2d_rb}
\end{align}
Note that
$\sum_{n_1=1}^{N_1}\sum_{n_2=1}^{N_2}P(n_1,n_2,t+1)=\sum_{n_1=1}^{N_1}\sum_{n_2=1}^{N_2}P(n_1,n_2,t)$,
which indicates that equations \eqref{eq:master_eq_2d},
\eqref{eq:master_eq_2d_lb} and \eqref{eq:master_eq_2d_rb} represent a
probability preserving Master equation.
\begin{figure}[!ht]
  \vspace{1em}
  \centering
  \begin{overpic}[tics=5, percent,width=0.5\textwidth]{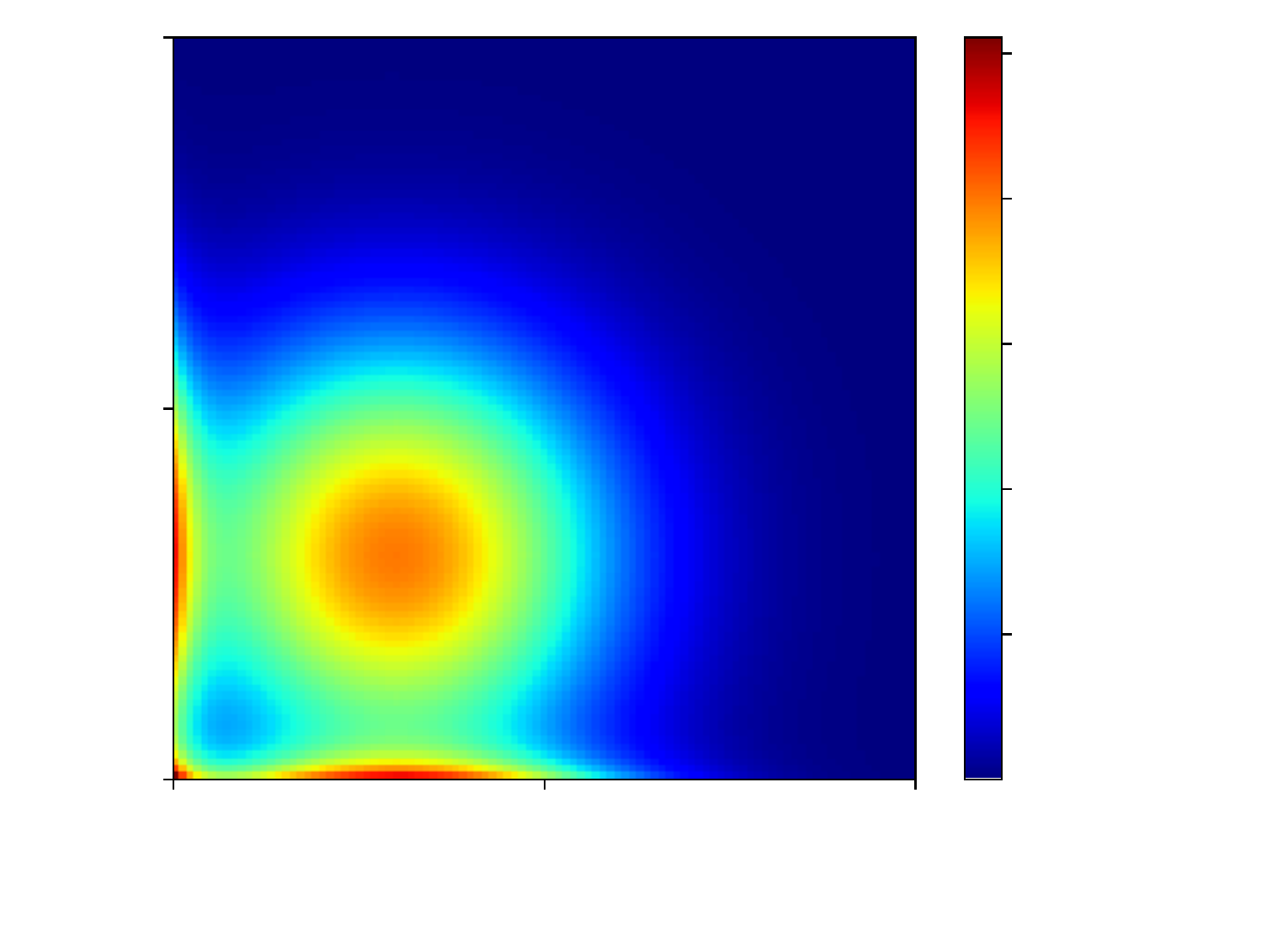}
    \fontsize{10}{10}
    \put(83.5, 66.5){$\times 10^{-4}$}
    \put(79.5, 66.5){$5$}
    \put(79.5, 55.1){$4$}
    \put(79.5, 43.75){$3$}
    \put(79.5, 32.35){$2$}
    \put(79.5, 21.15){$1$}
    \put(83.5, 31.35){\rotatebox{90}{$P_{\vec{n_0}}^{(r_1, r_2)}(n_1,
        n_2, t)$}}
    \put(40.2, 6){$51$}
    \put(68.2, 6){$101$}
    \put(12.5, 6){$1$}
    \put(10.1, 10){$1$}
    \put(8.2, 39.0){$51$}
    \put(6.3, 68){$101$}
    \put(3.2, 39.5){$n_2$}
    \put(40.0, 2.9){$n_1$}
  \end{overpic}
  \caption{(Colour Online) Two dimensional propagator, from
    \eeqref{eq:prop_2d_sum_t_r_1N}, evaluated at $t = 10^3$ with a
    domain of size $\vec{N} = (101, 101)$. The initial condition is
    $ \vec{n} =(71, 71) $, the diffusion parameters are
    $\vec{q} = (0.8, 0.8)$ and the bias parameters are
    $\vec{g} = (0.1, 0.1)$.  }
\label{fig:ref_2d} 
\end{figure}
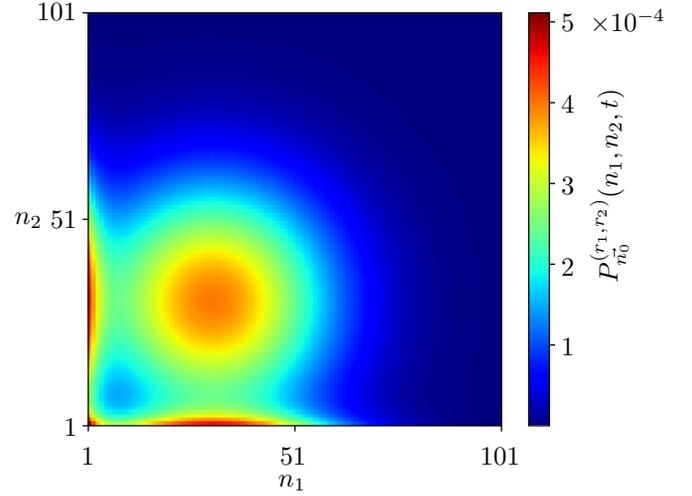

Symmetrising the dynamics and Fourier transforming along the second
dimension results in an effective 1D problem, analogous to
\eeqref{eq:matrix_master},
\begin{align}
\label{eq:q_master_effective}
\widehat{Q}\lb n_{_1}, \kappa_2, t+1\rb =  \sum_{\ell = 1}^{N_1} \mathbf{B}_{n_{_1}, \ell} \, \widehat{Q}\lb \ell,  \kappa_2, t \rb
\end{align}
where $\mathbf{B}$ is a tridiagonal matrix with elements on the upper
and lower diagonal being, respectively,
$\frac{\omega q_1}{2} \lb 1 + g_1 \rb $ and
$\frac{\omega q_1}{2} \lb 1 - g_1 \rb $ with
$\omega^{-1} = 1 - \frac{q_2}{2} + \frac{q_2}{2\eta_2}$. The elements
along the diagonal are
$B_{\ell, \ell} = \omega \ls 1 - \frac{q_1}{2} - \frac{q_2}{2} +
\frac{q_2}{\eta_2} \cos{\lb \kappa_2 \rb} \rs$, when
$\ell \neq 1, N_1$,
$B_{1, 1} = \omega \ls 1 - \frac{q_1}{4} \lb 1- g_1 \rb -
\frac{q_2}{2} + \frac{q_2}{\eta_2} \cos{\lb \kappa_2 \rb} \rs$,
$B_{N_1, N_1} = \omega \ls1 - \frac{q_1}{4} \lb 1+ g_1 \rb -
\frac{q_2}{2} + \frac{q_2}{\eta_2} \cos{\left(\kappa_2\right)}
\rs$. After supplementing the initial conditions
$\widehat{Q}(n_{_1}, \kappa_2, 0) = \delta_{n_{_1}, n_{0_1}}
\e^{-\mathrm{i} \kappa_2 n_{0_2}}f_2^{-\frac{n_{0_2}}{2}} \lb 1 -
f_2^{-1}\rb $, \eeqref{eq:q_master_effective}, due to the comparable
structure with \eeqref{eq:matrix_master}, can be solved explicitly in
$z$ and Fourier domains. Subsequently, inverse Fourier transforming
the second dimension, applying the method of images, reversing to the
asymmetric propagator and finally, after inverse $z$ transforming, one
obtains the exact spatio-temporal dependence (the calculation is
outlined in \apref{sec:two_prop_deriv}). Knowledge of the identity
\eeqref{eq:identity_ref}, allows us to write the time dependent
solution to the 2D random walks with independent bias in each
dimension as
\begin{align}
 &P^{(r_1, r_1)}_{\vec{n}_0}\lb n_{_1}, n_{_2}, t \rb = \lambda_1 \lambda _2  \nonumber \\
&+ \sum_{k_1 = 1}^{N_1- 1} \sum_{k_2 = 1}^{N_2 - 1} h_{k_1}^{\lb r_1 \rb}\lb n_{_1}, n_{0_1} \rb h_{k_2}^{\lb r_2 \rb}\lb n_{_2}, n_{0_2} \rb
\ls 1 +  \frac{s_{k_1}^{\lb r_1 \rb}}{2} +  \frac{s_{k_2}^{\lb r_2 \rb}}{2} \rs^t \nonumber \\
&\qquad + \lambda_1 \sum_{k_2 = 1}^{N_2 - 1} h_{k_2}^{\lb r_2 \rb}\lb n_{_2}, n_{0_2} \rb \ls 1+  \frac{s_{k_2}^{\lb r_2 \rb}}{2} \rs^t \nonumber \\
&\qquad + \lambda_2 \sum_{k_1 = 1}^{N_1- 1} h_{k_1}^{\lb r_1 \rb}\lb n_{_1}, n_{0_1} \rb \ls 1 +  \frac{s_{k_1}^{\lb r_1 \rb}}{2} \rs^t,
\label{eq:prop_2d_sum_t_r_1N}
\end{align}
where
$$
\lambda_i = \frac{f_i^{n_i - 1} \lb 1 - f_i \rb}{1 - f_i^{N_i}}.
$$
A similar procedure can be applied for the case of the absorbing,
periodic and mixed boundary conditions, although in the latter case, as metnioned earlier, the analytic solution is not fully explicit, but based on the
numerical roots of the orthogonal polynomials of the form
$f^{\frac{1}{2}}U_{s - 1}(x) - U_{s-2}(x)$.

In figure~\ref{fig:ref_2d}, we plot
$P^{(r_1, r_2)}_{\vec{n}_0}(n_1, n_2, t)$ for a specific time value
with the left-downward bias $\vec{g} = (0.1, 0.1)$. A feature worth
pointing out is the appearance of two saddle points that emerge at
intermediate times. They appear due to the steady-state probability at
the boundary being higher than the transient peak. In 1D this results
in the appearance of a local minimum.
\subsection{Propagator in Arbitrary dimensions and Arbitrary Boundary Conditions}
We use a hierarchical procedure to contruct bias lattice walk
propagators of any dimensions by generalising the procedure used to
derive the 2D random walk propagator \eeqref{eq:prop_2d_sum_t_r_1N} (the summary of the procedure can be found in
\apref{sec:higher_dims_construct_ap}). The resulting analytic
propagators are
\begin{align}
P^{(\vec{\gamma})}_{\vec{n}_0}(\vec{n}, t)\!&=\!\sum_{k_1 = w^{(\gamma_1)}}^{W^{(\gamma_1)}}\!\cdots\!
  \sum_{k_d = w^{(\gamma_d)}}^{W^{(\gamma_d)}}\ \prod_{j=1}^{d} h_{k_j}^{(\gamma_j)}(n_{_j}, n_{0_j})\nonumber \\
&\times\ls 1 + \frac{s_{k_1}^{(\gamma_1)}}{d}\!+\!\cdots\!+\!\frac{s_{k_d}^{(\gamma_d)}}{d} \rs^t,
\label{eq:full_nd_prop}
\end{align}
with $s^{(\gamma)}_k$ and $h^{(\gamma)}_k(n,n_0)$ defined,
respectively, in equations \eqref{eq:sk} and \eqref{eq:gk}, and with
$\omega^{(\gamma)}$ and $W^{(\gamma)}$ defined after
\eeqref{eq:1d_all_prop}. Using \eeqref{eq:full_nd_prop}, one can
derive first-passage (or first-return) probability and mean-first
passage times in higher dimensions with an abitrary combination of
reflecting and periodic boundaries which were previously unknown. Such
expressions enable one to study transport process that were, until
now, only possible through numerical means. In the following
subsections we employ \eeqref{eq:full_nd_prop} to reveal an intricate
bias depedence on the time dependent first-return probability, and we
study the effect of bias on the mean first-passage times in a
multi-target environment.
\begin{figure*}[hbt]
  \hfill
  \begin{overpic}[tics=2, percent,width=0.97\textwidth]{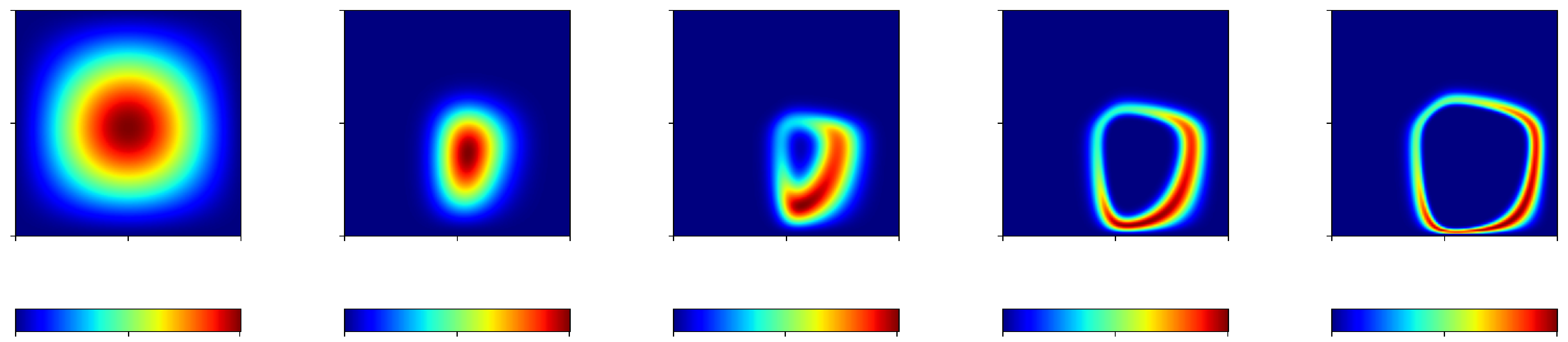}
    \fontsize{10}{10}
    \put(-2.7, 6.1){$-1$}
    \put(-1.1, 20.1){$1$}
    \put(-1.1, 13.1){$0$}
    \put(-3.5, 12.0){\rotatebox{90}{ $g_2$}}

    \put(7.7,  4.5){$0$}
    \put(-0.7, 4.5){$-1$}
    \put(14.8, 4.5){$1$}
    \put(6.6, 3.1){ $g_1$}
    \put(1.5, 18.5){\color{white}(a)}

    \put(28.7,  4.5){$0$}
    \put(20.3, 4.5){$-1$}
    \put(35.8, 4.5){$1$}
    \put(27.6,  3.1){ $g_1$}
    \put(22.5, 18.5){\color{white}(b)}

    \put(49.7,  4.5){$0$}
    \put(41.3, 4.5){$-1$}
    \put(56.8, 4.5){$1$}
    \put(48.6,  3.1){ $g_1$}
    \put(43.5, 18.5){\color{white}(c)}

    \put(70.7,  4.5){$0$}
    \put(62.3, 4.5){$-1$}
    \put(77.8, 4.5){$1$}
    \put(69.6,  3.1){ $g_1$}
    \put(64.5, 18.5){\color{white}(d)}

    \put(91.7,  4.5){$0$}
    \put(83.3, 4.5){$-1$}
    \put(98.8, 4.5){$1$}
    \put(90.6,  3.1){ $g_1$}
    \put(85.5, 18.5){\color{white}(e)}

    \put(-0.2, -1.7){$0.0$}
    \put(20.8, -1.7){$0.0$}
    \put(41.8, -1.7){$0.0$}
    \put(62.8, -1.7){$0.0$}
    \put(83.8, -1.7){$0.0$}

    \put(6.9, -1.7){$0.5$}
    \put(27.9, -1.7){$0.7$}
    \put(48.9, -1.7){$1.1$}
    \put(69.9, -1.7){$1.2$}
    \put(90.9, -1.7){$1.3$}

    \put(14.1, -1.7){$1.0$}
    \put(35.1, -1.7){$1.5$}
    \put(56.1, -1.7){$2.2$}
    \put(77.1, -1.7){$2.4$}
    \put(98.1, -1.7){$2.6$}

    \put(1.7,  -4.1){$R^{(\vec{r})}(\vec{n}, t) \ \times 10^{-2}$}
    \put(23.7, -4.1){$R^{(\vec{r})}(\vec{n}, t)\ \times 10^{-3}$}
    \put(43.7, -4.1){$R^{(\vec{r})}(\vec{n}, t)\ \times 10^{-4}$}
    \put(64.7, -4.1){$R^{(\vec{r})}(\vec{n}, t)\ \times 10^{-5}$}
    \put(85.7, -4.1){$R^{(\vec{r})}(\vec{n}, t)\ \times 10^{-6}$}
\end{overpic}
\vspace{30pt}
\caption{Return probability to the site
  $\vec{n} = (5, 18)$ as a function of the bias $\vec{g}$ for a 2D
  BLRW with reflecting boundaries. We use a domain of size
  $\vec{N} = (20, 20)$ and a diffusion parameter of value
  $\vec{q} = (0.8, 0.8)$. A positive (negative) $g_1$ indicates a
  drift to the left (right), while a positive (negative) $g_2$
  indicates a drift downwards (upwards). The panels (a) to (e)
  represent, respectively, $R^{(r)}(n, t)$ at time $t=10, 10^2, 10^3$,
  $10^4$ and $10^5$.}
\label{fig:return_p_2d} 
\end{figure*}
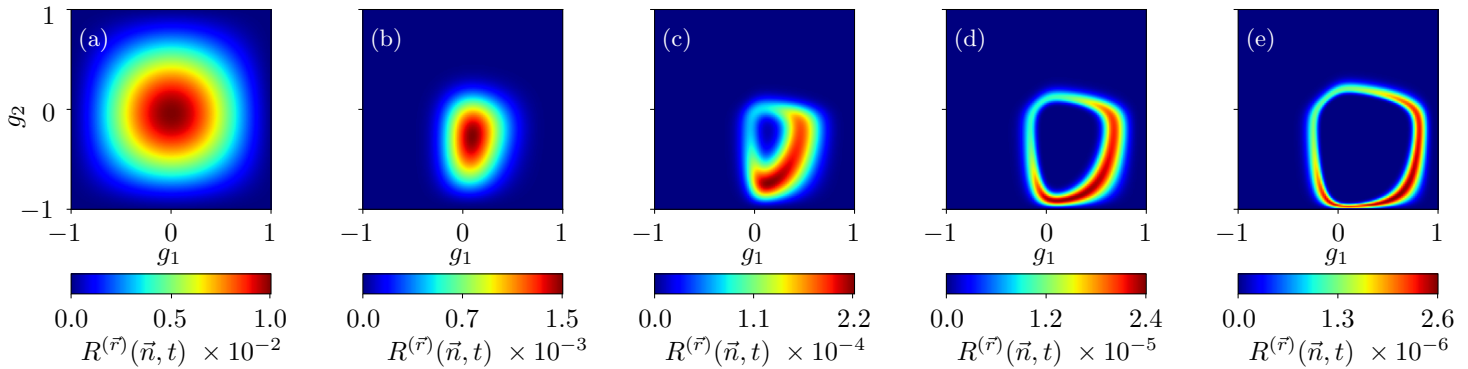
\subsubsection{First-Return Processes in Higher Dimensions}
A useful quantity in studying search processes is the probability of
the first recurrence of an event, that is the probability of a lattice
walker returning to the starting location for the first time. The
first-return probability, or henceforth, the return probability, is
derived via the renewal equation and in $z$-domain it is given by
$\wt{R}^{(\vec{r})}(\vec{n}, z) = 1 - \ls
\wt{P}^{(r)}_{\vec{n}}(\vec{n}, z) \rs^{-1}$. The study of the return
probability on lattice random walks has a long history
\cite{polya1919, polya1921}. Used originally for unbounded
$d$-dimensional lattices where it is found that a walker returns with
certainty to the starting location in 1D and 2D while for higher
dimensions there is a finite probability that the walker does not
return. Although the walker is bound to return to its initial position
in unbounded 1D and 2D domains, the mean return time (MRT) is always
infinite.  In bounded domains, on the other hand, the MRT is finite
and is equal to the reciprocal of the steady-state probability at the
site \cite{kac1947a}. For a LRW (without bias) the steady-state
probability is uniform and the MRT reduces to the domain size. In the
presence of a non uniform steady-state, as is the case with BLRW with
reflecting boundaries, the MRT, $\mathcal{R}^{(\vec{r})}_{\vec{n}}$,
becomes site-dependent and the return dynamics may be rather
complex. Namely, given an off-centre lattice site, one finds the MRT
to be minimised for a bias with a specific direction (see
\apref{sec:mean_return_ap}).

However, the MRT may hide the nuances of the temporal dynamics. In
order to examine the dynamics of the return probability, we use the
starting location $\vec{n} = (5, 18)$, and track the return
probability, $R^{(\vec{r})}(\vec{n}, t)$, at different times. We do
this in figure~\ref{fig:return_p_2d} by plotting
$R^{(\vec{r})}(\vec{n}, t)$ as a function $\vec{g}$. We use known
numerical methods \cite{abateetal1999} to invert the generating
function and plot in each panel the return probability for
progressively longer times from (a) to (e).

At short times (figure~\ref{fig:return_p_2d}a) one finds the return
probability to be independent of the bias direction as any bias pushes
the walker away from the starting location lowering the likelihood of
return. With $t=10^2$ in panel (b) we observe greater return
probabilities for certain values of $g_1 > 0 $ and $g_2 < 0 $. Since
the time $t$ is comparable to the shortest MRT (see
\apref{sec:mean_return_ap}), one expects the likelihood of returning at
$t = 10^2$ to be greater for the bias that yields the shortest MRT. A
further increase in time ($t=10^3$) results in the appearance of a
void. The void represents an area around a local minimum of
$R^{(\vec{r})}(\vec{n}, t)$. Its appearance indicates that a large
number of the trajectories for which the bias has values inside, have
already returned when compared to those with bias outside of the
void. Moreover, as the time scale is considerably larger than the one 
corresponding to the minimum of $\mathcal{R}^{(\vec{r})}_{\vec{n}}$,
there is no optimal bias to return. We thus observe an arched area of
high return probability in panel (c) compared to the area around a
maximum in panel (b). Increasing time further in panels (d) and (e)
results in the expansion of the void as stronger biases are necessary
to increase the probability of returning at longer times. One also
observes the radial stretching of the area of high return probability
as the dependence on the bias direction is progressively lost. In the
limit of large time the high values of $R^{(\vec{r})}(\vec{n}, t)$
acquires a square shape close to the extreme values of $\vec{g}$,
namely, $(-1,-1), (-1,1), (1,1)$ and $(1,-1)$.

\subsubsection{First-Passage Processes in Higher Dimensions}
Using the $z$ transform of \eeqref{eq:full_nd_prop} one can show that
the MFPT in higher dimensions with either reflective, periodic or a
mixture of the two types of boundaries, is given by
\begin{widetext}
\begin{align}
  T^{(\vec{\gamma})}_{\vec{n}_0 \to \vec{n}} = \frac{d}{\Omega}
  \underset{ k_1 + \cdots + k_d > 0}{\sum_{k_1 = 0}^{W^{(\gamma_1)}} \cdots\ \sum_{k_d = 0}^{W^{(\gamma_1)}}}
\frac{h_{k_1}^{(\gamma_1)}(n_{_1}, n_{0_d})\cdots h_{k_d}^{(\gamma_d)}(n_{_d}, n_{0_d})
-h_{k_1}^{(\gamma_1)}(n_{_1}, n_{_1})\cdots h_{k_1}^{(\gamma_d)}(n_{_d}, n_{_d})}
{s_{k_1}^{(\gamma_1)} + \cdots + s_{k_d}^{(\gamma_d)} }, 
\label{eq:full_mean_fpt}
\end{align}
\end{widetext}
where $\Omega = \prod_{j = 1}^d h_0^{(\gamma_j)}(n_{_j}, n_{0_j})$,
which is dependent only on $n_{_j}$ when $\gamma = r$, and is
independent of $n_{_j}$ and $n_{0_j}$ when $\gamma = p$; where
$s^{(\gamma)}_k$ and $h^{(\gamma)}_k(n,n_0)$ are defined,
respectively, in equation \eqref{eq:sk} and \eqref{eq:gk}; and with
$W^{(\gamma)}$ defined after \eeqref{eq:1d_all_prop}.

\begin{figure*}[ht]
  \hfill
  \begin{overpic}[tics=5,percent, width=0.9\textwidth]{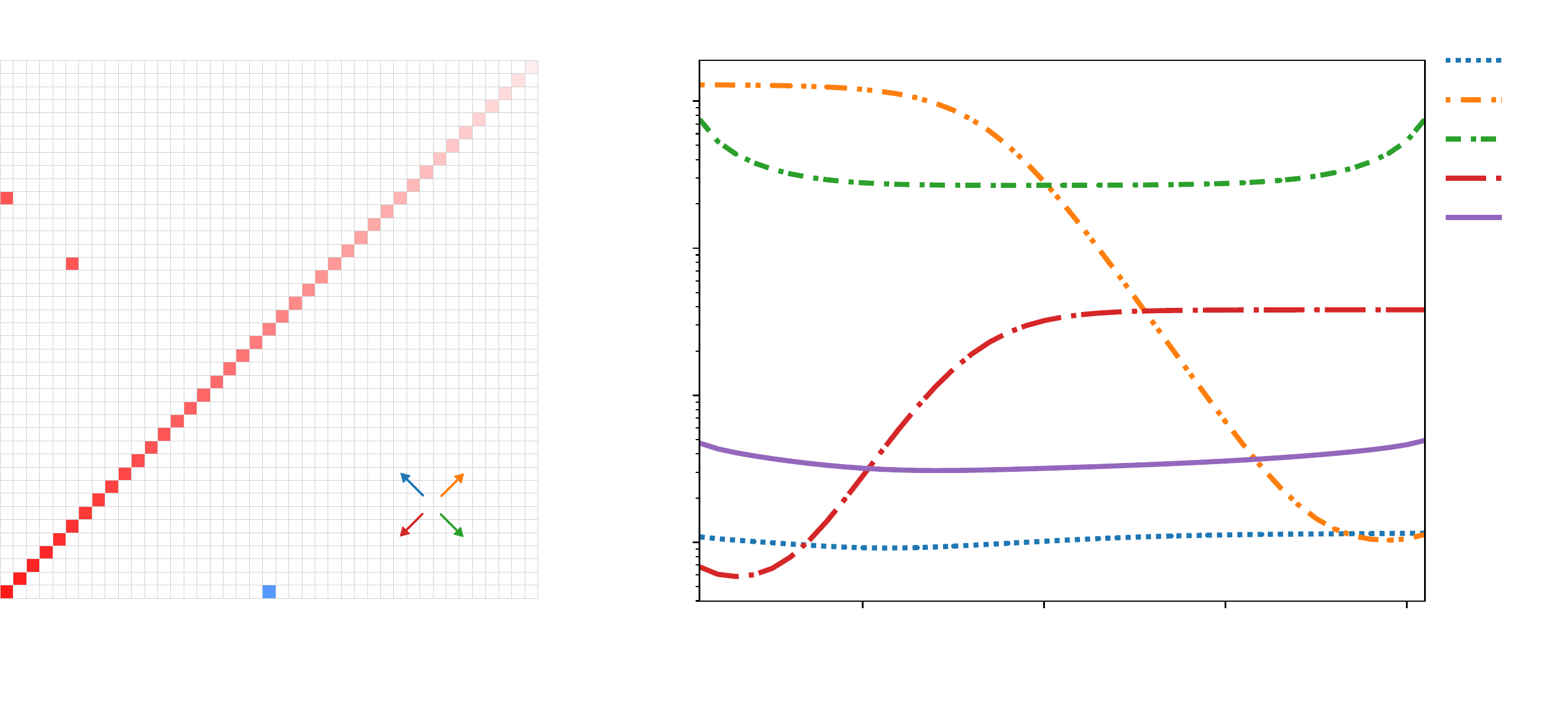}
     \put(96.5, 30.7){\fontsize{10}{10}  $\vec{g}_5$ }
     \put(96.5, 33.3){\fontsize{10}{10}  $\vec{g}_4$ }
     \put(96.5, 35.9){\fontsize{10}{10}  $\vec{g}_3$ }
     \put(96.5, 38.5){\fontsize{10}{10}  $\vec{g}_2$ }
    \put(96.5, 41.01){\fontsize{10}{10} $\vec{g}_1 $ }
    \put(40.7, 9.4){$10^3$}
    \put(40.7, 18.7){$10^4$}
    \put(40.7, 28){$10^5$}
    \put(40.7,  37.5){$10^6$}
    \put(54, 3.8){10}
    \put(65.5, 3.8){20}
    \put(77, 3.8){40}
    \put(88.5, 3.8){40}
    \put(0, 42.5){(a)}
    \put(45, 42.5){(b)}
    \put(37, 16){\rotatebox{90}{$T_{\vec{n}_0 \to {\{\vec{n}_1,
            \vec{n}_2, \vec{n}_2}\}}$}}
    \put(18, 8){\fontsize{10}{10}$\vec{n}_0$}
    \put(1.3, 33.2){\fontsize{10}{10}$\vec{n}_2$}
    \put(5.5, 29.0){\fontsize{10}{10}$\vec{n}_3$}
    \put(1, 12.0){\fontsize{10}{10}$\vec{n}_1$}
    \put(23, 15.7){\fontsize{10}{10}$\vec{g}_1$}
    \put(30.5, 15.7){\fontsize{10}{10}$\vec{g}_2$}
    \put(30.5, 8.5){\fontsize{10}{10}$\vec{g}_3$}
    \put(23, 8.5){\fontsize{10}{10}$\vec{g}_4$}

    \put(-2.0, 6.5){$1$}
    \put(-3.8, 40){$N_2$}
    \put(-0.3, 4){$1$}
    \put(32.8, 4){$N_1$}
    \put(68, 1){$m$}
  \end{overpic}
  \caption{(Colour Online) MFPT in a 2D domain of size $\vec{N} = (41, 41)$ with
    reflective boundary conditions to either of three targets. The
    walker is initially at $\vec{n}_0 = (21, 1)$ with diffusion
    parameter $\vec{q} = (0.8, 0.8)$. The coordinates of the first
    target are $\vec{n}_1 = (m, m)$ with $1 \leq m \leq 41 $, while
    that of the second and third targets are static with respective
    positions $\vec{n}_2 = (1, 31)$ and $\vec{n}_3 = (6, 26)$. The
    different biases considered are $\vec{g}_1 = (0.1, -0.1)$,
    $\vec{g}_2 = (-0.1, -0.1)$, $\vec{g}_3 = (-0.1, 0.1)$,
    $\vec{g}_4 = (0.1, 0.1)$ and the fully diffusive case
    $\vec{g}_5 = (0, 0)$. The panel (a) shows a schematic diagram of
    the setup with the positions of the targets, the initial condition
    and the directions of the different biases. In panel (b) for each
    of the bias we plot the MFPT as a function of the position of the
    third target.}
  \label{fig:mean_fpt_2d}
\end{figure*}
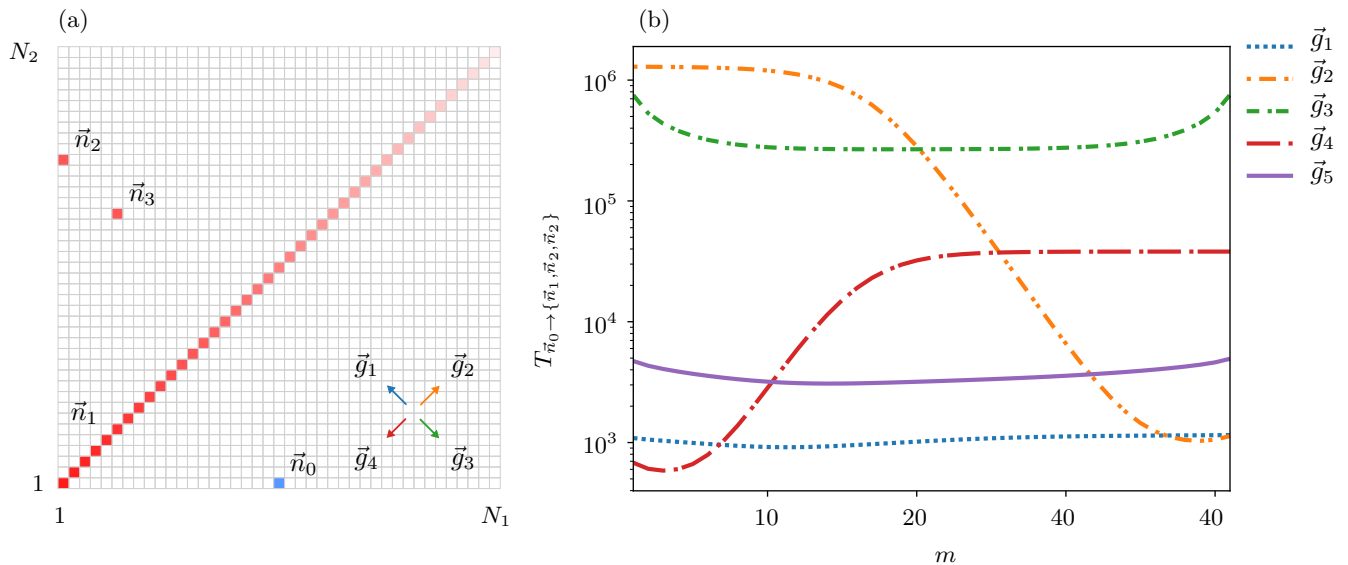

The first-passage dynamics becomes very rich in the presence of
multiple targets as the bias towards a specific target influences
dramatically the time it takes to reach either of the targets. We show
this dependence by plotting in figure~\ref{fig:mean_fpt_2d} the MFPT
to either of three targets as a function of the position of the first
target in a 2D box with reflective boundaries. We use
\eeqref{eq:full_mean_fpt} and the MFPT expression to either of three targets
from reference \cite{giuggioli2020}. Figure~\ref{fig:mean_fpt_2d}(a)
depicts the schematic diagram of the lattice, the biases and the
position of the targets. The position of the targets $\vec{n}_2$ and
$\vec{n}_3$ are fixed while the position of the first target
$\vec{n}_1 = (m, m)$ is slid along the diagonal.

The bias $\vec{g}_1$ shows the least dependence on the position of the
first target. With the bias $\vec{g}_1$, the walker always has a high
probability of reaching the second or third target regardless of the
position of the first. The shorter MFPT in this case occurs when the
line connecting $\vec{n}_0$ to $\vec{n}_1$ is parallel to $\vec{g}_1$,
i.e. when $m=11$. The diffusive case, $\vec{g}_5$, shows a slightly
stronger dependence on $m$ than $\vec{g}_1$, with its shortest MFPT
value being attributed to when the first target is closer to
$\vec{n}_0$, that is when $m = 14$. As the direction of the bias
$\vec{g}_3$ is always away from the targets, the MFPT decreases as the
first target is moved. It gives the minimum MFPT in correspondence to
the shortest distance that the walker travels against the bias to
reach $\vec{n}_1$.

With $\vec{g}_2$ and $\vec{g}_4$ being opposite to each other and
parallel to any of the positions of the first target, the
corresponding MFPT displays similar characteristics. When $m = 1$
($m = 41$), corresponding with the first target located in the
bottom-left (top-right) corner, the bias $\vec{g}_4$ ($\vec{g}_2$)
exhibits the shortest MFPT due to the bias pushing trajectories
towards the corner. As $m$ is increased from $m = 1$, the MFPT of
$\vec{g}_4$ increases due to $\vec{n}_1$ moving out of the bottom-left
corner. Analogously, as $m$ is decreased from $m = 41$, $\vec{n}_1$
moves out of the top-right corner causing the MFPT of $\vec{g}_2$ to
increase.  The difference in the high values of the MFPT of
$\vec{g}_4$ when $m > 30$, and the MFPT of $\vec{g}_2$ when $m < 10$,
is due to the position of the second and third targets. With
$\vec{n}_2$ and $\vec{n}_3$ being closer to the bottom-left than the
top-right corner, one expects the largest MFPT of $\vec{g}_4$ to be
smaller than the largest MFPT of $\vec{g}_2$, and vice versa for the
shortest MFPT.

An interesting observation is that when the first target is positioned
in the top-right corner with $m > 35$, the MFPT of $\vec{g}_2$ and
$\vec{g}_1$ are comparable. It highlights the strong dependence of the
MFPT on the positioning of the targets relative to the boundary
corner. In the presence of a bias, one can achieve shorter or similar
MFPTs by positioning fewer targets close to the corner and in the
direction of the bias ($\vec{g}_2$ case with high $m$) as opposed to
many targets away from it ($\vec{g}_1$ case).

Our observations are particularly relevent in the domain of
field-driven translocation in channels with periodic
corrugation. Here, one is interested in the first-passage times of
tracer particle moving under an external bias. Recent numerical
analysis \cite{valovetal2020}, reveal that particles travel close to
the boundaries as they pass through a funnel. For further work, it
would be interesting to reaffirm such results, by studying the MFPT
using a similar setup to figure~\ref{fig:mean_fpt_2d}a, but with
targets concentrated in the corner and by changing the initial
position instead of a target position. With the two orthogonal
boundaries acting like a funnel one expects similar results to those
observed numerically.

\section{Conclusions}
\label{sec:conclusion}
We conclude, by reminding that while the continuous-time BLRW in
confined domains has been studied extensively in 1D
\cite{khanthabalakrishnan1984, khanthabalakrishnan1985a}, a thorough
treatment of the analogous discrete time case was missing from the
literature and only the propagator for the case $q = 1$ and with
absorbing boundaries was known \cite{godoyfujita1992}. Compared to
reference \cite{godoyfujita1992}, here we have derived the generating
function of the 1D propagator in finite domains by employing an
alternative procedure yielding both finite series and compact
expressions, with the latter used to create the generating function
for the first-passage probability in reflecting and periodic
domains. In order to find the time-dependent propagators, we have used
known results for tridiagonal matrices with perturbed corners
\cite{yuehcheng2006, willms2008} instead of inverting the generating
function of the propagators via a contour integral. From the finite
series time-dependent propagators we have recovered the known
solutions to the drift-diffusion equation, linking the movement
parameters of a BLRW with the drift velocity and the diffusion
coefficient for a Brownian walker.

By exploiting the properties of Chebyshev polynomials, the generating
function of the first-passage probability with periodic and reflective
boundaries was inverted explicitly to yield the exact time
dependence. Surprisingly, the periodic case was shown to display
bimodal features when its analog without bias is known to be
monomodal. Lastly, by employing a hierarchical dimensional reduction
we have derived time-dependent propagators for the confined BLRW 
in any number of dimensions and with arbitrary boundary conditions. The propagators
were then used to find explicit expressions for the mean first-passage
time in a $d$-dimensional box, torus or an arbitrary combination of
both. The generating function of the propagators have also highlighted
the influence a bias may have on the time dependence of the return
probability.

\begin{acknowledgments}
  LG acknowledges funding the Engineering and Physical Research
  Council Grant (EPSRC) nos.
  BB/T012196/1 and EP/I013717/1, while SS acknowledges funding from
  EPSRC Grant no. S108151-111.
\end{acknowledgments}
\bibliography{full_library.bib} 

\providecommand{\noopsort}[1]{}
\begin{thebibliography}{10}

\bibitem{schillingpartzschbook2012}
R.~L. Schilling and L.~Partzsch, {\em Brownian Motion: An Introduction to
  Stochastic Processes}.
\newblock De {{Gruyter}} Graduate, {Berlin ; Boston}: {De Gruyter}, 1st ed~ed.,
  2012.

\bibitem{hughesbook1995}
B.~D. Hughes, {\em Random Walks and Random Environments}.
\newblock {Clarendon Press; Oxford University Press}, 1995.

\bibitem{fisher1966}
M.~E. Fisher, ``Shape of a {{Self}}-{{Avoiding Walk}} or {{Polymer Chain}},''
  {\em The Journal of Chemical Physics}, vol.~44, pp.~616--622, 1966.

\bibitem{godrecheetal2017}
C.~Godr{\`e}che, S.~N. Majumdar, and G.~Schehr, ``Record statistics of a
  strongly correlated time series: Random walks and {{L\'evy}} flights,'' {\em
  Journal of Physics A: Mathematical and Theoretical}, vol.~50, p.~333001,
  2017.

\bibitem{ewensbook2004}
W.~J. Ewens, {\em Mathematical {{Population Genetics}}: {{I}}. {{Theoretical
  Introduction}}}.
\newblock {Springer New York}, 2004.

\bibitem{okubolevinbook2001}
A.~Okubo and S.~A. Levin, {\em Diffusion and Ecological Problems: Modern
  Perspectives}.
\newblock {New York, USA}: {Springer Verlag}, second~ed., 2001.

\bibitem{lohmarkrug2009}
I.~Lohmar and J.~Krug, ``Diffusion-limited reactions and mortal random walkers
  in confined geometries,'' {\em Journal of Statistical Physics}, vol.~134,
  pp.~307--336, 2009.

\bibitem{zumofenblumen1982}
G.~Zumofen and A.~Blumen, ``Energy transfer as a random walk. {{II}}.
  {{Two}}-dimensional regular lattices,'' {\em The Journal of Chemical
  Physics}, vol.~76, pp.~3713--3731, 1982.

\bibitem{pearlstein1982}
R.~M. Pearlstein, ``Exciton {{Migration}} and {{Trapping}} in
  {{Photosynthesis}},'' {\em Photochemistry and Photobiology}, vol.~35, no.~6,
  pp.~835--844, 1982.

\bibitem{mayoetal2011}
M.~L. Mayo, E.~J. Perkins, and P.~Ghosh, ``First-passage time analysis of a
  one-dimensional diffusion-reaction model: Application to protein transport
  along {{DNA}},'' {\em BMC Bioinformatics}, vol.~12, p.~S18, 2011.

\bibitem{shinkolomeisky2019}
J.~Shin and A.~B. Kolomeisky, ``Target search on {{DNA}} by interacting
  molecules: {{First}}-passage approach,'' {\em The Journal of Chemical
  Physics}, vol.~151, no.~12, p.~125101, 2019.

\bibitem{maierbrockmann2017}
B.~F. Maier and D.~Brockmann, ``Cover time for random walks on arbitrary
  complex networks,'' {\em Physical Review E}, vol.~96, p.~042307, 2017.

\bibitem{grassberger2017}
P.~Grassberger, ``How fast does a random walk cover a torus?,'' {\em Physical
  Review E}, vol.~96, p.~012115, 2017.

\bibitem{riascosetal2020}
A.~P. Riascos, D.~Boyer, P.~Herringer, and J.~L. Mateos, ``Random walks on
  networks with stochastic resetting,'' {\em Physical Review E}, vol.~101,
  p.~062147, 2020.

\bibitem{thielsokolov2016}
F.~Thiel and I.~M. Sokolov, ``Effective-medium approximation for lattice random
  walks with long-range jumps,'' {\em Physical Review E}, vol.~94, p.~012135,
  2016.

\bibitem{rednerbook2001}
S.~Redner, {\em A Guide to First-Passage Processes}.
\newblock {Cambridge, UK}: {Cambridge University Press}, 2001.

\bibitem{metzleretalbook2014}
R.~Metzler, G.~Oshanin, and S.~Redner, {\em First-Passage Phenomena and Their
  Applications}.
\newblock {World Scientific}, 2014.

\bibitem{benichouvoituriez2014}
O.~B{\'e}nichou and R.~Voituriez, ``From first-passage times of random walks in
  confinement to geometry-controlled kinetics,'' {\em Physics Reports},
  vol.~539, pp.~225--284, 2014.

\bibitem{kac1947a}
M.~Kac, ``On the notion of recurrence in discrete stochastic processes,'' {\em
  Bull. Amer. Math. Soc.}, vol.~53, pp.~1002--1010, 1947.

\bibitem{condaminetal2005a}
S.~Condamin, O.~B{\'e}nichou, and M.~Moreau, ``First-{{Passage Times}} for
  {{Random Walks}} in {{Bounded Domains}},'' {\em Physical Review Letters},
  vol.~95, p.~260601, 2005.

\bibitem{condaminetal2007}
S.~Condamin, V.~Tejedor, and O.~B{\'e}nichou, ``Occupation times of random
  walks in confined geometries: {{From}} random trap model to diffusion-limited
  reactions,'' {\em Physical Review E}, vol.~76, p.~050102, 2007.

\bibitem{giuggioli2020}
L.~Giuggioli, ``Exact {{Spatiotemporal Dynamics}} of {{Confined Lattice Random
  Walks}} in {{Arbitrary Dimensions}}: {{A Century}} after {{Smoluchowski}} and
  {{P\'olya}},'' {\em Physical Review X}, vol.~10, p.~021045, 2020.

\bibitem{hughesbookvol1_1995}
B.~D. Hughes, {\em Random walks and random environments: random walks}, vol.~1.
\newblock Oxford: Clarendon Press, 1995.

\bibitem{hughesbookvol2_1995}
B.~D. Hughes, {\em Random walks and random environments: random walks}, vol.~2.
\newblock Oxford: Clarendon Press, 1995.

\bibitem{sourjikwingreen2012}
V.~Sourjik and N.~S. Wingreen, ``Responding to chemical gradients: Bacterial
  chemotaxis,'' {\em Current Opinion in Cell Biology}, vol.~24, pp.~262--268,
  2012.

\bibitem{bergbook1993}
H.~C. Berg, {\em Random Walks in Biology}.
\newblock {Princeton, N.J}: {Princeton University Press}, expanded ed~ed.,
  1993.

\bibitem{jekely2009}
G.~J{\'e}kely, ``Evolution of phototaxis,'' {\em Philosophical Transactions of
  the Royal Society B: Biological Sciences}, vol.~364, pp.~2795--2808, 2009.

\bibitem{sweietal2018}
O.~Swei, J.~Gregory, and R.~Kirchain, ``Does {{Pavement Degradation Follow}} a
  {{Random Walk}} with {{Drift}}? {{Evidence}} from {{Variance Ratio Tests}}
  for {{Pavement Roughness}},'' {\em Journal of Infrastructure Systems},
  vol.~24, p.~04018027, 2018.

\bibitem{hillhader1997}
N.~A. Hill and D.~P. H{\"a}der, ``A {{Biased Random Walk Model}} for the
  {{Trajectories}} of {{Swimming Micro}}-organisms,'' {\em Journal of
  Theoretical Biology}, vol.~186, pp.~503--526, 1997.

\bibitem{mabroukietal2007}
I.~Mabrouki, G.~Froc, and X.~Lagrange, ``Biased {{Random Walk Model}} to
  {{Estimate Routing Performance}} in {{Sensor Networks}},'' in {\em 9\`eme
  {{Rencontres Francophones}} Sur Les {{Aspects Algorithmiques}} Des
  {{T\'el\'ecommunications}}}, pp.~73--76, 2007.

\bibitem{valovetal2020}
A.~Valov, V.~Avetisov, S.~Nechaev, and G.~Oshanin, ``Field-driven tracer
  diffusion through curved bottlenecks: Fine structure of first passage
  events,'' {\em Physical Chemistry Chemical Physics}, vol.~22, no.~33,
  pp.~18414--18422, 2020.

\bibitem{godoyfujita1992}
S.~Godoy and S.~Fujita, ``Reflection principles for biased correlated walks.
  {{Simple}} applications,'' {\em Journal of mathematical physics}, vol.~33,
  no.~9, pp.~2998--3003, 1992.

\bibitem{montroll1967}
E.~W. Montroll, ``Stochastic processes and chemical kinetics,'' in {\em
  Energetics in Metallurgic Phenomena: Vol {{III}}.} (W.~Mueller, ed.),
  pp.~122--187, {New York}: {Gordon and Breach}, 1967.

\bibitem{giuggiolietal2019}
L.~Giuggioli, S.~Gupta, and M.~Chase, ``Comparison of two models of tethered
  motion,'' {\em Journal of Physics A: Mathematical and Theoretical}, vol.~52,
  p.~075001, 2019.

\bibitem{khanthabalakrishnan1984}
M.~Khantha and V.~Balakrishnan, ``Hopping conductivity of a one-dimensional
  bond-percolation model in a constant field: {{Exact}} solution,'' {\em
  Physical Review B}, vol.~29, pp.~4679--4690, 1984.

\bibitem{khanthabalakrishnan1985a}
M.~Khantha and V.~Balakrishnan, ``Reflection principles for biased random walks
  and application to escape time distributions,'' {\em Journal of statistical
  physics}, vol.~41, no.~5-6, pp.~811--824, 1985.

\bibitem{yueh2005}
W.-C. Yueh, ``Eigenvalues of several tridiagonal matrices,'' {\em Appl. Math.
  e-notes}, vol.~5, no.~66-74, pp.~210--230, 2005.

\bibitem{yuehcheng2008}
W.-C. Yueh and S.~S. Cheng, ``Explicit eigenvalues and inverses of tridiagonal
  {{Toeplitz}} matrices with four perturbed corners,'' {\em The ANZIAM
  Journal}, vol.~49, p.~361, 2008.

\bibitem{willms2008}
A.~R. Willms, ``Analytic {{Results}} for the {{Eigenvalues}} of {{Certain
  Tridiagonal Matrices}},'' {\em SIAM Journal on Matrix Analysis and
  Applications}, vol.~30, pp.~639--656, 2008.

\bibitem{montrollweiss1965}
E.~W. Montroll and G.~H. Weiss, ``Random walks on lattices. {{II}},'' {\em
  Journal of Mathematical Physics}, vol.~6, pp.~167--181, 1965.

\bibitem{polya1919}
G.~P{\'o}lya, ``Quelques problemes de probabilit\'e se rapportant a la
  promenade au hasard?,'' {\em L'Enseignement Math\'ematique}, vol.~20,
  no.~444, p.~8, 1919.

\bibitem{polya1921}
G.~P{\'o}lya, ``\"uber eine {{Aufgabe}} der {{Wahrscheinlichkeitsrechnung}}
  betreffend die {{Irrfahrt}} im {{Stra\ss ennetz}},'' {\em Mathematische
  Annalen}, vol.~84, no.~1-2, pp.~149--160, 1921.

\bibitem{abateetal1999}
J.~Abate, G.~L. Choudhury, and W.~Whitt, ``An introduction to numerical
  transform inversion and its application to probability models,'' in {\em
  Computational probability} (W.~Grassman, ed.), pp.~257--323, Boston: Kluwer,
  1999.

\bibitem{yuehcheng2006}
W.-C. Yueh and S.~S. Cheng, ``Explicit eigenvalues and inverses of several
  {{Toeplitz}} matrices,'' {\em The ANZIAM Journal}, vol.~48, pp.~73--97, 2006.

\bibitem{lhopitalbook2015}
G.~F. A.~d. L'Hôpital and J.~Bernoulli, {\em L'{Hôpital}'s analyse des
  infiniments petits: an annotated translation with source material by {Johann}
  {Bernoulli}}.
\newblock No.~volume 50 in Science networks historical studies, Cham:
  Birkhäuser, 2015.
\newblock OCLC: 911263365.

\bibitem{montrollwest1979}
E.~W. Montroll and B.~J. West, ``On an enriched collection of stochastic
  processes,'' in {\em Studies in Statistical Mechanics: Vol {{VII}}.
  {{Fluctuation}} Phenomena} (E.~Montroll and J.~Lebowitz, eds.), pp.~61--175,
  {Amsterdam}: {North Holland Publishing}, 1979.

\bibitem{grigolini2006}
P.~Grigolini, ``The continuous time random walk versus the generalized master
  equation,'' {\em Advances in Chemical Physics}, vol.~133, pp.~357--474, 2006.

\bibitem{polyaninbook2002}
A.~D. Polyanin, {\em Handbook of Linear Partial Differential Equations for
  Engineers and Scientists}.
\newblock {Boca Raton}: {Chapman \& Hall/CRC}, 2002.

\end{thebibliography}
\bibliographystyle{ieeetr.bst}

\begin{widetext}
\appendix
\counterwithin{figure}{section}
\section{Derivation of Propagators in 1D with Reflective Boundaries}
\subsection{Single Reflective Boundary}
\label{sec:symmterisation_single}
The Kronecker delta initial condition for the propagator
$\mathcal{P}^{(r)}_{n_0}(n, 0) = \delta_{n, n_0}$ gives an initial
condition for the symmetric propagator with $\mu = f^{-\frac{1}{2}}$
in \eeqref{eq:q_transform} equal to
$\mathcal{Q}^{(a)}(n, 0) = f^{-\frac{n}{2}} \delta_{n, n_0} -
f^{-\frac{n + 2}{2}} \delta_{n+1, n_0}$. Convoluting this initial
condition with the symmetric propagator \eqref{eq:hfunc} and
accounting for the contribution of the image of the initial condition
via
$\wt{\mathcal{Q}}^{(a)}(n, z) = \sum_{m = 0}^{\infty}
\mathcal{Q}^{(a)}(m, 0)\ls \wt{H}_{m}(n, z) - \wt{H}_{-m}(n, z)
\big.\rs$ gives
\begin{equation}
\label{eq:q_bounded_a_0}
 \wt{\mathcal{Q}}^{(a)}(n , z)  = 
 \frac{\vphi^{-\left| n - n_0 \right|}
   - \vphi^{-\left| n + n_0 \right|} }
  {\ls 1 - z \omega \lb 1 - q \rb \rs \sqrt{1 - \zeta^2}} 
-
 \frac{f^{-\frac{1}{2}} \lb \vphi^{-\left| n - n_0 + 1 \right|}
   - \vphi^{-\left| n + n_0 - 1\right|} \rb}
  {\ls 1 - z \omega \lb 1 - q \rb \rs \sqrt{1 - \zeta^2}},
\end{equation}
where $\zeta$ and $\vphi$ are defined in \eeqref{eq:zeta} and
\eeqref{eq:phi} respectively. Using \eeqref{eq:q_bounded_a_0} in
\eeqref{eq:inverse_q_transform} yields
\begin{equation}
  \label{eq:inverse_q_bounded_a_0}
  \wt{\mathcal{P}}^{(r)}_{n_0}(n , z)  = 
  \frac{1}
  {\ls 1 - z \lb 1 - q \rb \rs \sqrt{1 - \lb\frac{\beta}{\eta}\rb^2}} 
  \sum_{j = 0}^{\infty}f^{-\frac{j}{2}}
  \lb
  \alpha^{-\left| n - n_0  + j\right|}
  - \alpha^{-\left| n + n_0  + j\right|} \rb
  -
  f^{-\frac{j+ 1}{2}} \lb \alpha^{-\left| n - n_0 + j + 1 \right|}
  - \alpha^{-\left| n + n_0 + j- 1\right|} \rb.
\end{equation}
To assist in evaluating \eeqref{eq:inverse_q_bounded_a_0}, it is useful
to consider the full summation as differences of two series. The
difference involving $\alpha^{-\left| n - n_0 \right|}$ terms
produces
\begin{equation}
  \label{}
  \sum_{j = 0}^{\infty} f^{-\frac{j}{2}} \alpha^{-\left| n - n_0  + j\right|} -
  \sum_{j = 0}^{\infty}f^{-\frac{j+1}{2}} \alpha^{-\left| n  - n_0 + j + 1\right|} 
  = \alpha^{-\left| n - n_0 \right|},
\end{equation}
as the only surviving term is the $j = 0$ term,
whereas evaluating the sum with the terms
$\alpha^{-\left| n + n_0 \right|}$ gives
\begin{equation}
  \label{}
\sum_{j = 0}^{\infty} f^{-\frac{j + 1}{2}} \alpha^{-\left| n + n_0  + j - 1\right|} -
\sum_{j = 0}^{\infty}f^{-\frac{j}{2}} \alpha^{-\left| n + n_0 + j \right|}
=
\frac{\alpha^{-\left| n + n_0 \right|} \left( \alpha - f^{\frac{1}{2}} \right)}{f^{\frac{1}{2}} - \alpha^{-1}}.
\end{equation}
Hence the propagator with a single reflective boundary
between the sites $n = 0$ and $n = 1$ in $z$-domain is 
\begin{equation}
  \label{}
\wt{\mathcal{P}}_{n_0}^{(r)}(n, z) = 
  \frac{1}{\ls 1 - z \lb 1 - q \rb \rs \sqrt{1 - \zeta^2}} \left( \alpha^{-\left| n - n_0 \right|}  +
\frac{\alpha^{-\left| n + n_0 \right|} \left( \alpha - f^{\frac{1}{2}} \right)}{f^{\frac{1}{2}} - \alpha^{-1}} \right).
\end{equation}
With some simple algebra one then obtains \eeqref{eq:p_bounded_r_1} in
the main text.

\subsection{Two Reflective Boundaries}
\label{sec:symmetrisation_two_bound}
With the domain being finite, one must construct the bounded
propagator with an infinite number of images of the propagator
\eqref{eq:hfunc} with shifted initial conditions. Similar to the case
with a single reflecting boundary, the initial condition
$P^{(r)}_{n_0}(n, 0) = \delta_{n, n_0}$ translates into
$\mu = f^{-\frac{1}{2}}$ and
$Q^{(a)}(n, 0) = f^{-\frac{n}{2}} \delta_{n, n_0} - f^{-\frac{n +
    2}{2}}\delta_{n+1, n_0}$. Convolution of this initial condition
with the general solution constructed with infinite images of the
unbounded propagator \eqref{eq:hfunc} via
$\wt{Q}^{(a)}(n, z) = \sum_{m = 0}^{N} \sum_{k = -\infty}^{+\infty}
Q^{(a)}(m , 0) \ls \big. \wt{H}_{m + 2kN}(n, z) - \wt{H}_{-m + 2kN}(n,
z) \rs $ gives
\begin{equation}
  \label{}
\wt{Q}^{(a)}(n, z) =
\sum_{k = -\infty}^{\infty}
 \frac{\vphi^{-\left| n - n_0 + 2kN\right|}
   - \vphi^{-\left| n + n_0 +  2kN\right|} }
  {\ls 1 - z \omega \lb 1 - q \rb \rs \sqrt{1 - \zeta^2}} 
-
f^{-\frac{1}{2}}\sum_{k = -\infty}^{\infty}
 \frac{\vphi^{-\left| n - n_0 + 1+  2kN\right|}
   - \vphi^{-\left| n + n_0  - 1 +   2kN\right|} }
  {\ls 1 - z \omega \lb 1 - q \rb \rs \sqrt{1 - \zeta^2}}.
\end{equation}
Summing the images yields the solution to the symmetric propagator
with absorbing boundaries at $n = 0$ and $n = N$
\begin{align}
\label{eq:q_bounded_a_0N}
\wt{Q}^{(a)}(n, z) &= 
\frac{\vphi^{N - \lp n - n_0 \rp} + \vphi^{-N + \lp n - n_0 \rp} - \vphi^{N - \lp n + n_0 \rp}  -\vphi^{-N + \lp n + n_0 \rp}}
{\lb \vphi^N - \vphi^{-N} \rb \ls 1 - z \omega \lb 1 - q \rb \rs \sqrt{1 - \zeta^2}} \nonumber \\
&-
f^{-\frac{1}{2}}\lc \frac{\vphi^{N - \lp n - n_0 + 1\rp} + \vphi^{-N + \lp n - n_0 + 1\rp} -
\vphi^{N - \lp n + n_0 - 1\rp}  -\vphi^{-N + \lp n + n_0 -1\rp}}
{\lb \vphi^N - \vphi^{-N} \rb \ls 1 - z \omega \lb 1 - q \rb \rs \sqrt{1 - \zeta^2} } \rc.
\end{align}

To transform back to the symmetric propagator we apply the
transformation \eqref{eq:inverse_q_transform} to
\eeqref{eq:q_bounded_a_0N} and we obtain
\begin{align}
  \label{eq:inverse_q_bounded_a_0N}
 \wt{\mathcal{P}}^{(r)}_{n_0}(n , z)  &= 
\sum_{j = 0}^{\infty}f^{-\frac{j}{2}} 
\lc \frac{\alpha^{N - \lp n - n_0 \rp} + \alpha^{-N + \lp n - n_0 \rp} - \alpha^{N - \lp n + n_0 \rp}  -\alpha^{-N + \lp n + n_0 \rp}}
{\lb \alpha^N - \alpha^{-N} \rb \ls 1 - z  \lb 1 - q \rb \rs \sqrt{1 - \lb \frac{\eta}{\beta}\rb^2}}
                                         \rc \nonumber  \\
&- \sum_{j = 0}^{\infty}f^{-\frac{j+1}{2}} 
\lc \frac{\alpha^{N - \lp n - n_0 + 1\rp} + \alpha^{-N + \lp n - n_0 + 1\rp} -
\alpha^{N - \lp n + n_0 - 1\rp}  -\alpha^{-N + \lp n + n_0 -1\rp}}
{\lb \alpha^N - \alpha^{-N} \rb \ls 1 - z  \lb 1 - q \rb \rs \sqrt{1 - \lb \frac{\eta}{\beta}\rb^2} } \rc.
\end{align}

We proceed in a similar fashion as before by considering pairwise
differences of the series. The
$\alpha^{\pm N \mp \left| n - n_0 \right|}$ terms result in the
difference of two geometric series, namely,
\begin{equation}
  \label{}
\alpha^{\pm N \mp \left| n - n_0 \right|} = 
\sum_{j = 0}^{\infty} f^{-\frac{j}{2}} \alpha^{\pm N \mp \left| n - n_0  + j\right|} -
\sum_{j = 0}^{\infty}f^{-\frac{j+1}{2}} \alpha^{\pm N \mp \left| n  - n_0 + j + 1\right|},
\end{equation}
while the $\alpha^{\pm N \mp \left| n + n_0 \right|}$ terms
produce a difference of the following geometric series
\begin{equation}
  \label{}
\alpha^{\pm N \mp \left| n + n_0 \right|} \ls \frac{ \alpha - f^{\frac{1}{2}} }{f^{\frac{1}{2}} - \alpha^{-1}} \rs^{\pm 1}
=
\sum_{j = 0}^{\infty} f^{-\frac{j + 1}{2}} \alpha^{\pm N \mp \left| n + n_0  + j - 1\right|} -
\sum_{j = 0}^{\infty}f^{-\frac{j}{2}} \alpha^{\pm N \mp \left| n + n_0 + j \right|},
\end{equation}
Putting everything together we find
\begin{equation}
\wt{P}^{(r)}_{n_0}(n, z) = 
\frac{\eta f^{\frac{n - n_0}{2}}}{z q \sinh\ls \acosh \lb \frac{\eta}{\beta} \rb \rs}
\lc \frac{\alpha^{N - \lp n - n_0 \rp} + \alpha^{-N + \lp n - n_0 \rp} - \alpha^{N - \lp n + n_0 \rp} \xi  -\alpha^{-N + \lp n + n_0 \rp} \xi^{-1}}
{2\sinh\ls N \acosh{\lb \frac{\eta}{\beta} \rb} \rs} \rc, 
\end{equation}
and with some further algebra we obtain \eeqref{eq:p_bounded_r_1N} in
the main text.

\section{Time Dependent Solution with Mixed Boundary Condition}
\label{sec:time_dep_mixed}
We rewrite the mixed propagator \eqref{eq:p_bounded_m} in terms of
Chebyshev polynomials of the second kind,
\begin{equation}
  \wt{P}_{n_0}^{(m)} (n, z) = 
  \frac{2 \eta f^{\frac{n-n_0}{2}}  U_{N - n_> - 1} \lb \frac{\eta}{\beta} \rb
    \lc f^{\frac{1}{2}} U_{n_< - 1} \lb \frac{\eta}{\beta} \rb -
      U_{n_< - 2} \lb \frac{\eta}{\beta} \rb \rc}
  {z q \lc f^{\frac{1}{2}} U_{N - 1} \lb \frac{\eta}{\beta} \rb -
      U_{N - 2} \lb \frac{\eta}{\beta} \rb \rc } 
\end{equation}
To find the inverse $z$ transform of \eeqref{eq:p_bounded_m}, we first
find the roots of the orthogonal polynomial
$f^{\frac{1}{2}} U_{N-1}\lb \sigma \rb - U_{N-2}\lb \sigma
\rb$. Defining $\cos\lb \theta_k \rb$ as the roots, the time dependent
solution is then written as
\begin{equation} 
  \label{}
2 f^{\frac{n - n_0}{2}} 
  \sum_{k = 1}^{N-1} \lim_{\sigma \to \cos\lb \theta_k \rb} \
  \frac{\ls 1 - p + \frac{p}{\eta} \cos\lb\theta_k \rb \rs^t
    \ls \big. \sigma - \cos\lb \theta_k \rb \rs
    U_{N - n_> - 1}\lb \sigma \rb
    \lc  f^{\frac{1}{2}} U_{n_< - 1}\lb \sigma \rb - U_{n_< - 2}\lb \sigma \rb\rc} 
  {f^{\frac{1}{2}} U_{N - 1}\lb \sigma \rb - U_{N - 2}\lb \sigma \rb}.
\end{equation}

To evaluate the limit we apply L'Hôpital's rule
\cite{lhopitalbook2015} to obtain
\begin{equation}
  \label{}
  2 f^{\frac{n - n_0}{2}} 
  \sum_{k = 1}^{N-1}
  \frac{\ls 1 - p + \frac{p}{\eta} \cos\lb\theta_k \rb \rs^t
    U_{N - n_> - 1}\lb \sigma \rb
    \lc  f^{\frac{1}{2}} U_{n_< - 1}\lb \sigma \rb - U_{n_< - 2}\lb \sigma \rb\rc} 
  {- \lim_{\sigma \to \cos\lb \theta_k \rb} \frac{1}{\sin^2\lb \theta_k \rb}
    \lc N \ls f^{\frac{1}{2}} T_N(\sigma) - T_{N-1} (\sigma)\rs + T_{N-1} -\cos(\theta_k) \ls f^{\frac{1}{2}} U_{N - 1}\lb \sigma \rb - U_{N - 2}\lb \sigma \rb \rs\rc},
\end{equation}
and with some further algebraic manipulation we obtain
\eeqref{eq:1d_all_prop} with
$\gamma = m$.

\section{Continuous Time and Spatial Limits of the One Dimensional Propagators}
\label{sec:continuous_limit_ap}
The continuous space and time propagators of the biased lattice walk
in finite domains, that is the drift-diffusion bounded propagator can
be recovered by appropriate limiting procedures. We consider first the
continuous-time discrete-space analog of the 1D propagators
(i.e. \eeqref{eq:1d_all_prop} in the main text) given by 
$C^{(\gamma)}_{n_0}(n, \tau) = \sum_{s = 0}^{\infty} W(s, \tau)
P_{n_0}^{(\gamma)}(n, s)$ \cite{montrollwest1979, grigolini2006}, where
$W(s, \tau)$ is the probability of $s$ jumps to occur in (continuous)
time $\tau$. With $\psi{\lb \tau \rb}$, the probability of a jump
event to occur at time $\tau$, one can construct
$\overline{W}(s, \epsilon) = \frac{1 -
  \overline{\psi}(\epsilon)}{\epsilon} \overline{\psi}{(\epsilon)}$,
where
$\overline{f}(\epsilon) = \int_{0}^{\infty}\e^{-\epsilon t} f{(t)}
\mathrm{d}t $ is the Laplace transform of $f(t)$. Laplace transforming
and evaluating the geometric sum yields
\begin{align}
  \label{eq:1d_cont_lap}
  \overline{C}^{(\gamma)}_{n_0}(n, \epsilon) = 
  \sum_{k = w^{(\gamma)}}^{W^{(\gamma)}}
  \frac{ h_{k}^{(\gamma)}\lb n, n_0 \rb \ls 1 - \overline{\psi}(\epsilon) \rs }
  {\epsilon \lc 1 - \overline{\psi}(\epsilon) \ls 1 + s_k^{(\gamma)}\rs \rc}.
\end{align}
Defining $\psi(\tau) = 2R\e^{-2R\tau}$ where $R$ is a rate, and its
Laplace transform $\overline{\psi}(\epsilon) =2R/(\epsilon + 2 R)$,
one can inverse Laplace transform \eeqref{eq:1d_cont_lap} obtaining
the continuous-time discrete-space biased random walk in finite domains 
\begin{align}
C_{n_0}^{(\gamma)}(n, \tau) = 
\sum_{k = w^{(\gamma)}}^{W^{(\gamma)}}
h_{k}^{(\gamma)}\lb n, n_0 \rb \Huge{\e}^{2R\tau s_k^{(\gamma)}}
\label{eq:1d_cont_time}
\end{align}
To take the continuous spatial limit of \eeqref{eq:1d_cont_time}, we
consider a lattice spacing $b$ with $b, g \to 0$ and
$R, N, n, n_0 \to +\infty$, such that $x = bn$, $x_0 = b n_0$,
$L = Nb$, $Rqb^2 \to D$, and $\sfrac{g}{b} \to \sfrac{v}{2 D}$, where
$L$ is the domain size ($0 \leq x, x_0 \leq L$), $D$ the diffusion
constant and $v$ the drift velocity. Evaluation of these limits
requires different steps for each of the boundary conditions which are
outlined in the following sections.
\subsection{Absorbing Boundaries}
From \eeqref{eq:1d_cont_time}, the continuous-time discrete-space
propagator with absorbing boundaries is given by
\begin{align}
C_{n_0}^{(a)}(n, \tau) &= 
\sum_{k=1}^{N-1} \frac{ 2 f^{\frac{n - n_0}{2}}
    \sin\ls \left(\frac{n-1}{N-1}\right)k\pi\rs
    \sin\ls \left(\frac{n_0-1}{N-1}\right)k\pi\rs}{N-2}\nonumber \\
&\times \exp{\lc -2Rq\tau \ls 1 - \frac{1}{\eta} \cos{\lb \frac{k \pi}{N-1}\rb}\rs \rc}.
\label{eq:1d_cont_time_ab}
\end{align}
The term $f^{\frac{n - n_0}{2}}$ in \eeqref{eq:1d_cont_time_ab} needs to
be rewritten as $\exp{\ls \frac{1}{2} \lb n - n_0\rb \ln\lb f \rb\rs}$
before Taylor expansion of $\ln\lb f\rb$ with
$f = \frac{1 - g}{1 + g}$ to obtain
\begin{equation}
\exp{\ls \frac{1}{2} \lb n - n_0\rb \ln\lb f \rb\rs}
\to
\exp{\ls \frac{ v \lb x_0 - x \rb }{2 D}\rs},
\end{equation}
where $bn \to x$, $bn_0 \to x_0$ and $g/b \to v/\lb 2D \rb$.  The time
dependent term in \eeqref{eq:1d_cont_time} requires a Taylor expansion
of the $\cos(\theta)$ term around $\theta = 0$ and
$\eta = \lb \sqrt{1 - g^2} \rb^{-1}$ around $g = 0$. 
With $Nb \to L$, one then has
\begin{equation}
2 R p\ls 1 - \frac{1}{\eta}\cos{\lb \frac{k \pi}{N - 1} \rb} \rs \to
\frac{v^2}{4D} + \frac{D \pi^2 k^2}{L^2}.
\end{equation}
Combining these results we recover the continuous space-time solution
to the drift-diffusion equation with absorbing boundaries (see
e.g. equation (1.1.4-7) in reference \cite{polyaninbook2002}),
\begin{align}
  \label{eq:1d_cont_time_space_ab}
  C^{(a)}_{x_0}(x, \tau) &= \frac{2}{L}
  \sum_{k=1}^{\infty} \sin\lb \frac{k \pi x}{L} \rb \sin\lb \frac{k \pi x_0}{L} \rb \nonumber \\
&\times \exp{\ls  -\frac{D\pi^2 k^2 \tau}{L^2}  + \frac{2 v \lb x_0 - x \rb - \tau v^2 }{4D}\rs}.
\end{align}
\subsection{Reflecting Boundaries}
Using \eeqref{eq:1d_cont_time}, the continuous-time discrete-space
propagator with two reflective boundaries is
\begin{align}
\label{eq:1d_cont_time_ref}
 C_{n_0}^{(r)}(n, \tau) &= 
\frac{f^{n - 1} \left( 1 - f \right)}{1 - f^N} + 
\frac{2f^{\frac{n - n_0}{2}}}{N} \sum_{k=1}^{N-1}
\frac{ \lb f^{\frac{1}{2}} \sin \ls \frac{n k \pi}{N} \rs - \sin \ls \lb n - 1\rb
\frac{k \pi }{N}\rs \rb \lb f^{\frac{1}{2}} \sin \ls \frac{n_0 k \pi}{N} \rs - \sin \ls \lb n_0 - 1\rb
\frac{k \pi }{N}\rs \rb}{ 1 + f   - 2 f^{\frac{1}{2}}\cos\lb \frac{k \pi}{N} \rb}
\nonumber \\
&\times \exp{\lc -2Rp\tau \ls 1 - \frac{1}{\eta} \cos{\lb \frac{k \pi}{N}\rb}\rs \rc}. 
\end{align}
For the continuous spatial limit the procedure is analogous to the
case with two absorbing boundaries. The important differences with the absorbing case are the
steady-state term,
\begin{equation}
  \lim_{\tau \to\infty}
  C_{n_0}^{(r)}(n, \tau) =
  \frac{f^{n - 1} \lb 1 - f \rb}{1 - f^N},
\end{equation}
and the
$\frac{1}{2} \lb 1 + f \rb - f^{\frac{1}{2}}\cos\lb \frac{k \pi}{N}
\rb$ term in \eeqref{eq:1d_cont_time_ref}. Starting with the steady
state term and rewriting
\begin{equation}
  \label{}
  \frac{f^{n - 1} \lb 1 - f \rb}{1 - f^{N}} =
  \frac{(1 - f) \exp{\ls (n - 1)\ln\lb f \rb \rs}}{1 - \exp{\ls N \ln{\lb f\rb} \rs}}, 
\end{equation}
one has to expand
the $\ln\lb f\rb $ term, as done for the absorbing case, before taking the limits. The steady-state
probability density in the continuous case becomes
\begin{equation}
  \label{}
\lim_{t\to \infty} C_{x_0}(x, \tau) = \frac{v\exp{\lb \frac{v x}{D} \rb}}{D \ls 1 -  \exp{\lb \frac{v L}{D} \rb} \rs}.
\end{equation}
For the terms inside the summation, it is convenient to expand the $n$
and $n_0$ dependence first and rewrite
\begin{align}
  \label{}
 &\frac{ \lb f^{\frac{1}{2}} \sin \ls \frac{n k \pi}{N} \rs - \sin \ls \lb n - 1\rb
\frac{k \pi }{N}\rs \rb \lb f^{\frac{1}{2}} \sin \ls \frac{n_0 k \pi}{N} \rs - \sin \ls \lb n_0 - 1\rb
\frac{k \pi }{N}\rs \rb}{ 1 + f   - 2 f^{\frac{1}{2}}\cos\lb \frac{k \pi}{N} \rb} 
\nonumber \\
&= \frac{\times \lc \sin \ls \frac{n k \pi}{N} \rs \lb f^{\frac{1}{2}}\csc{\ls \frac{k\pi}{N} \rs}  - \cot{\ls \frac{k\pi}{N} \rs}\rb + 
\cos \ls \frac{ n k\pi }{N} \rs \rc
\lc \sin \ls \frac{ n_0 k  \pi }{N} \rs \lb f^{\frac{1}{2}}\csc{\ls \frac{k\pi}{N} \rs}  - \cot{\ls \frac{k\pi}{N} \rs}\rb + 
\cos \ls \frac{n_0 k \pi  }{N} \rs \rc}{(1 + f)\csc^{2}{\ls \frac{k\pi}{N} \rs} + 2f^{\frac{1}{2}} \cot{\ls \frac{k\pi}{N} \rs} \csc{\ls \frac{k\pi}{N} \rs}}.
\end{align}
With a Taylor expansion of
$f^{\frac{1}{2}}\csc{\ls \theta \rs} - \cot{\ls \theta
  \rs}$,  around $\theta = 0$, 
where $\theta = \frac{k \pi}{N}$ is the expansion variable,
we find the limits
\begin{equation}
  \label{}
\mathlarger{\lim}_{\substack{N \to \infty \\ b,g \to 0}}
\lc f^{\frac{1}{2}}\csc{\ls \frac{k\pi}{N} \rs} - \cot{\ls \frac{k\pi}{N} 
  \rs} \rc
=
-\frac{vL}{2 D k \pi}, 
\end{equation}
similarly,

\begin{equation}
  \label{}
\mathlarger{\lim}_{\substack{N \to \infty \\ b,g \to 0}}
\lc  (1 + f)\csc^{2}{\ls \frac{k\pi}{N} \rs} + 2f^{\frac{1}{2}} \cot{\ls \frac{k\pi}{N} \rs} \csc{\ls \frac{k\pi}{N} \rs}  \rc
=
\frac{v^2L^2}{4 D^2 k^2 \pi^2}. 
\end{equation}
We finally recover the continuous space-time propagator with two
reflective boundaries
\begin{align}
 C_{x_0}^{(r)}(x, \tau) &= 
\frac{v\exp{\lb \frac{v x}{D} \rb}}{D \ls 1 -  \exp{\lb \frac{v L}{D} \rb} \rs}
+ \frac{2}{L}\exp{\ls \frac{2 v \lb x_0 - x \rb - \tau v^2 }{4D} \rs}\nonumber \\
&\times \mathlarger{\sum}_{k=1}^{\infty}\frac{
\lc \cos \ls \frac{k \pi x  }{L} \rs  -
\mu_k \sin \ls \frac{k  \pi x }{L} \rs \rc
\lc \cos \ls \frac{k \pi x_0  }{L} \rs  -
\mu_k \sin \ls \frac{k  \pi x_0 }{L} \rs \rc}
{\lb 1 + \mu_k^2 \rb}e^{- \frac{D\pi^2 k^2 \tau}{L^2}},
\label{eq:1d_cont_time_space_re}
\end{align}
where $\mu_k = vL/(2Dk\pi)$, (e.g. see (1.1.4-8) in reference\cite{polyaninbook2002}).

\subsection{Mixed Boundaries}
The discrete-space continuous-time propagator with mixed boundary
conditions is 
\begin{align}
  \label{eq:1d_cont_time_m}
C_{n_0}^{(m)}(n, \tau) &= 
\frac{2f^{\frac{n - n_0}{2}}}{N-1} \sum_{k=1}^{N-1}
\frac{2f^{\frac{n - n_0}{2}}\sin \ls \lb N - n_> \rb \theta_k \rs \lc  f^{\frac{1}{2}} \sin \ls n_< \theta_k  \rs -
  \sin \ls \lb n_<  - 1\rb \theta_k \rs \rc} {
    \lb N - 1 \rb \cos\ls \lb N-1\rb \theta_k\rs - N f^{\frac{1}{2}} \cos{\ls N \theta_k \rs}} \nonumber \\
&\times \exp{\lc -2Rq\tau \ls 1 - \frac{1}{\eta} \cos{\lb \frac{k \pi}{N-1}\rb}\rs \rc}.
\end{align}
Before taking the limits on the spatial dependence, it is necessary to
study first the effect of the limits on the relationship
\begin{align}
  \label{}
f^{\frac{1}{2}} U_{N-1}(\sigma) - U_{N-2}(\sigma) = 0.
\label{eq:ortho_rel}
\end{align}
We rewrite the Chebyshev polynomials in \eeqref{eq:ortho_rel} using
their trigonometric definition to yield
\begin{align}
  \label{}
f^{\frac{1}{2}} \sin{\ls N \acos{\lb \sigma \rb} \rs} - \sin{\ls (N-1)  \acos{\lb \sigma \rb} \rs} = 0.
\end{align}
Expanding the $\sin\ls \lb N - 1 \rb \acos\lb \sigma \rb \rs$ results in the relationship

\begin{equation}
  \label{eq:tanmixrel}
\tan{\ls N \acos{\lb \sigma \rb} \rs}  
  = \frac{\sin{\ls \acos{\lb \sigma \rb} \rs}}
{ \cos\ls \acos\lb \sigma \rb\rs - f^{\frac{1}{2}} }.
\end{equation}

Defining $\theta_k = b \rho_k = \acos{\lb \sigma \rb}$ and
substituting we find
\begin{equation}
  \label{}
  \frac{\tan{\lb L \rho_k \rb}}{\rho_k \cos{\lb b \rho_k \rb}} =
  \frac{\sin{\lb b \rho_k \rb}}{\rho_k\ls \cos^2{\lb b \rho_k \rb} -  f^{\frac{1}{2}} \cos{\lb b \rho_k \rb}\rs},
\end{equation}
which, in the limit $b \to 0 $ and $ g \to 0$, results in the
transcendental equation
\begin{equation}
  \label{eq:trans_root}
\frac{\tan{\lb L \rho_k \rb}}{\rho_k} = \frac{2D}{v}.
\end{equation}

Expanding the numerator and denominator in the spatial dependence
$h_{k}^{(m)}(n, n_0)$, and with the help of \eeqref{eq:tanmixrel} we find

\begin{equation}
  \smaller
h_k^{(m)}(n, n_0) = 
\frac{2f^{\frac{n - n_0}{2}}
  \lb \cos{\lb \theta_k\rb - f^{\frac{1}{2}}} \rb
  \lc \bigg. \tan\lb N \theta_k \rb \cos\lb  n_< \theta_k\rb -\sin\lb n_>  \theta_k \rb \rc
  \lc \bigg.  \tan{\lb N \theta_k\rb} \cos \lb n_< \theta_k\rb
-\sin \lb   n_<  \theta_k\rb \rc}{
 \lb N - 1 \rb \ls \cos{\lb \theta_k \rb} 
+ \tan{\lb N \theta_k \rb} \sin\lb \theta_k\rb \rs- N f^{\frac{1}{2}} }.
\label{eq:h_k_m_limit_tan}
\end{equation}
With the $n_>$ dependence being
equivalent to the $n_<$ dependence in \eeqref{eq:h_k_m_limit_tan}, we rewrite
\begin{equation}
h_k^{(m)}(n, n_0) = 
\frac{f^{\frac{n - n_0}{2}}}{B_k} \ls
\bigg.
\tan\lb N \theta_k \rb \cos\lb n  \theta_k \rb -\sin\lb n  \theta_k \rb \rs
\ls
\bigg.
\tan{\lb N \theta_k\rb} \cos \lb  n_0  \theta_k \rb
-\sin \lb n_0 \theta_k \rb \rs  
\end{equation}
where
$$
B_k=  \frac{
 \lb N - 1 \rb \ls \cos{\lb \theta_k \rb} 
+ \tan{\lb N \theta_k \rb} \sin\lb \theta_k\rb \rs- N f^{\frac{1}{2}} }
{2 \ls \cos{\lb \theta_k\rb - f^{\frac{1}{2}}} \rs}.
$$

After substituting $b\rho_k = \theta_k$, in the continuous limit we
find
\begin{equation}
h_k^{(m)}(x, x_0) = 
\frac{1}{A_k} \exp{\ls \frac{v (x_0 - x)}{2D} \rs} \ls
\bigg.
\tan\lb L \rho_k \rb \cos\lb x \rho_k \rb -\sin\lb x  \rho_k \rb \rs
  \ls \bigg. \tan{\lb L \rho_k\rb} \cos \lb x_0 \rho_k  \rb
-\sin \lb x_0  \rho_k \rb \rs  
\end{equation}
where
\begin{equation}
A_k = \lim_{b \to 0} b  B_k=  \lim_{b \to 0} \lc \frac{
 \lb L - b \rb \ls \cos{\lb b \rho_k \rb} 
+ \tan{\lb L \rho_k \rb} \sin\lb b \rho_k\rb \rs- L f^{\frac{1}{2}} }
{2 \ls \cos{\lb b \rho_k\rb - f^{\frac{1}{2}}} \rs} \rc
=
- \frac{2 D}{v} + \frac{L}{\cos^{2}{\left(L \rho_k \right)}}.
\end{equation}
Putting everything together we recover the continuous-time
continuous-space solution (see Equation~(1.1.4-9) in
\cite{polyaninbook2002}) with mixed boundary conditions
\begin{align}
 C_{x_0}^{(m)}(x, \tau) &= 
\exp{\ls \frac{2 v \lb x_0 - x \rb - \tau v^2 }{4D} \rs}
\mathlarger{\sum}_{k=1}^{\infty}
\frac{1}{A_k}
\ls\bigg.
\tan\lb L \rho_k \rb \cos\lb x \rho_k \rb -\sin\lb x  \rho_k \rb \rs \nonumber \\
 &\times \ls \bigg. \tan{\lb L \rho_k\rb} \cos \lb x_0 \rho_k  \rb
-\sin \lb x_0  \rho_k \rb \rs ,
\label{eq:1d_cont_time_space_mix}
\end{align}
where $\rho_k$ are the roots of \eeqref{eq:trans_root}.
\subsection{Periodic Boundaries}
Starting with the discrete-space continuous-time propagator
\begin{align}
  \label{eq:1d_cont_time_p}
C_{n_0}^{(p)}(n, \tau) = 
\frac{1}{N} \sum_{k=0}^{N-1}
  \exp{\lc \frac{2 k \pi \mathrm{i} \lb n - n_0 \rb}{N}  -2Rp\tau
  \ls 1 - \cos{\lb \frac{2 k \pi}{N}\rb} - \mathrm{i} g \sin{\lb \frac{2 k \pi}{N}\rb} \rs \rc},
\end{align}
one needs to expand the $\cos$ and $\sin$ terms before taking the
limits resulting in the following continuous space-time propagator
\begin{align}
  \label{eq:1d_cont_time_space_p}
C_{x_0}^{(p)}(x, \tau) = 
\frac{1}{L} \sum_{k=0}^{\infty}
  \exp{\lc \frac{2 k \pi \mathrm{i} \lb x - x_0 + v\rb}{L} 
   -\frac{4 Dk^2 \pi^2 \tau}{L^2}  \rc},
\end{align}
which clearly satisfies the periodic boundary condition. For higher
dimensions, the limiting procedure can be carried through to give the
continuous space-time analog which is the product of the
one dimensional propagators (equations
\eqref{eq:1d_cont_time_space_ab}, \eqref{eq:1d_cont_time_space_re},
\eqref{eq:1d_cont_time_space_mix} and \eqref{eq:1d_cont_time_space_p})
along each direction.
\section{First-Passage Probability and Related Quantities with
  Reflecting and Periodic One Dimensional Domain}
\label{sec:first_passage_ap}
The first-passage probability, being the ratio of propagators in the
$z$ domain, can be constructed, respectively, for reflecting and
periodic domains from \eeqref{eq:p_bounded_r_1N} and
\eeqref{eq:p_bounded_p} yielding
\begin{equation}
  \wt{F}^{(r)}_{n_0}(n, z) =
  \frac{\wt{P}^{(r)}_{n_0}\lb n, z \rb}{\wt{P}^{(r)}_{n}\lb n, z \rb} = \left\{
\begin{array}{cr}
  \frac{f^{\frac{n - n_0}{2}} \left( f^{\frac{1}{2}} \sinh{\ls n_0\, \acosh{\lb \frac{\eta}{\beta} \rb}\rs}
  -\sinh{\ls \lb  n_0 - 1\rb \, \acosh{\lb \frac{\eta}{\beta} \rb}\rs}\right)}
  {f^{\frac{1}{2}} \sinh{\ls n\, \acosh{\lb \frac{\eta}{\beta} \rb}\rs}
  -\sinh{\ls \lb  n - 1\rb \, \acosh{\lb \frac{\eta}{\beta} \rb}\rs}}, &n > n_0 \\ [\bigskipamount]
  \frac{f^{\frac{n - n_0}{2}} \left( f^{\frac{1}{2}} \sinh{\ls \lb N - n_0 \rb \acosh{\lb \frac{\eta}{\beta} \rb}\rs}
  -\sinh{\ls \lb N + 1 -  n_0 \rb \, \acosh{\lb \frac{\eta}{\beta} \rb}\rs}\right)}
  {f^{\frac{1}{2}} \sinh{\ls \lb N- n \rb\, \acosh{\lb \frac{\eta}{\beta} \rb}\rs}
  -\sinh{\ls \lb N + 1-  n\rb \, \acosh{\lb \frac{\eta}{\beta} \rb}\rs}}, &n < n_0 ,
\end{array} \right.
\label{eq:fpz_1d_r}
\end{equation}
and 
\begin{equation}
  \wt{F}^{(p)}_{n_0}(n, z) =
  \frac{\wt{P}^{(p)}_{n_0}\lb n, z \rb}{\wt{P}^{(p)}_{n}\lb n, z \rb} = \left\{
\begin{array}{cr}
  \frac{f^{\frac{n - n_0}{2}} \lc f^{\frac{1}{2}} \sinh{\ls \lb N - n +n_0 \rb \acosh{\lb \frac{\eta}{\beta} \rb}\rs}
  + f^{-\frac{N}{2}} \sinh{\ls \lb n -  n_0 \rb \acosh{\lb \frac{\eta}{\beta} \rb }\rs } \rc}
  {\sinh \ls N \acosh{\lb \frac{\eta}{\beta} \rb}\rs}, &n < n_0 \\ [\bigskipamount]
  \frac{f^{\frac{n - n_0}{2}} \lc f^{\frac{1}{2}} \sinh{\ls \lb N - n_0 +n \rb \acosh{\lb \frac{\eta}{\beta} \rb}\rs}
  + f^{\frac{N}{2}}\sinh{\ls \lb n_0 -  n \rb \, \acosh{\lb \frac{\eta}{\beta} \rb}\rs}\rc}
  {\sinh\ls N \acosh{\lb \frac{\eta}{\beta} \rb}\rs}, &n > n_0 .
\end{array} \right.
\label{eq:fpz_1d_p}
\end{equation}
Using equations \eqref{eq:fpz_1d_r} and \eqref{eq:fpz_1d_p} yields the MFPT expressions in
\eqref{eq:mean_fpt_1d_r} and \eqref{eq:mean_fpt_1d_p} in the
main text.

From the generating function of the return probability,
$\wt{R}^{(r)}(n, z) = 1 - \ls \wt{P}_{n}(n, z) \rs^{-1}$, one can
compute the MRT
\begin{equation}
  \label{eq:mean_ret_1d}
  \mathcal{R}^{(r)}_n = \frac{1 - f^N}{f^{n - 1}\lb 1- f \rb}
  \quad \text{and}  \quad
  \mathcal{R}^{(p)}_n = N.
\end{equation}
which is, as expected from Kac's theorem, the reciprocal of
the steady-state probability at the site.

\subsection{First-Passage Probability when $g \pm 1$ and $q \neq 1$
  with Periodic Boundary}
\label{sec:fpp_periodic_limit}
Restricting the walker to move only forward or to remain at a site, that
is no back-tracking, involves taking the limits $g \to \pm 1$,
alternatively, $f \to 0 $ or $f\to \infty$ in the first-passage probability
expressions for the periodic domain. Although it is trivial to find
the periodic propagator in \eeqref{eq:1d_all_prop} for such limits, the
same cannot be said of the first-passage probability in \eeqref{eq:fpt_p}
due to the term $f^{\frac{n - n_0}{2}}$. In the latter case it is much
easier to derive the first-passage probability by separating the cases
$n < n_0$ and $n > n_0$, and constructing the expression
combinatorially. One can show that when $g = 1$ and $n < n_0$, or when
$g = -1$ and $n > n_0$, the first-passage probability becomes
\begin{equation}
  \label{}
F_{n_0}^{(p)}(n, t) = {t -1 \choose t -  \lp n - n_0 \rp } q^{\lp n - n_0 \rp}  \lb 1 - q \rb^{t - \lp n- n_0 \rp} 
\Theta\ls t - \lp n - n_0 \rp \big.\rs,
\label{eq:fpp_p_nb_m}
\end{equation}
while when $g = 1$ and $n > n_0$, or when $g = -1$ and $n < n_0$, it
is
\begin{equation}
  \label{}
  F_{n_0}^{(p)}(n, t) = {t -1 \choose t -  \lb N - \lp  n - n_0 \rp \rb }
  q^{ N - \lp n - n_0\rp} \lb 1 - q \rb^{t - \lb N -  \lp n - n_0 \rp \rb} 
\Theta\ls t - \lb N - \lp  n - n_0 \rp \rb \big.\rs.
\label{eq:fpp_p_nb_M}
\end{equation}

When \eeqref{eq:fpp_p_nb_m} applies, the probability of reaching the site is zero
if the number of time steps is smaller than the displacement between
the target and initial site, i.e. when $t < \lp n - n_0 \rp$. At
$t = \lp n - n_0 \rp $, the relation $F_{n_0}^{(p)}(n, t) = q^{\lp n - n_0\rp}$
indicates that the walker may reach the target by always moving with each step giving a contribution equal to $q$ to the probability. When
$t > \lp n - n_0 \rp$, the walker has a choice of remaining at any of
the sites between $n$ and $n_0$ (excluding $n$). In that case the coefficient
${t -1 \choose t - \lp n - n_0 \rp }$ represents all the possible combinations with which
the walker can reach $n$ from $n_0$ by making $|n-n_0|$ steps and $t-|n-n_0|$ pauses along the way. On the other
hand when \eeqref{eq:fpp_p_nb_M} applies, the walker
travels around the domain with the length of the shortest possible
path from $n$ to $n_0$ being $N - \lp n - n_0\rp$, and the meaning of the terms are analogous to the case in \eqref{eq:fpp_p_nb_m}. The case $q = 1$
corresponds to a Kronecker delta in time:
$F_{n_0}^{(p)}(n, t) = \delta_{t, \lp n - n_0\rp} $ or
$F_{n_0}^{(p)}(n, t) = \delta_{t, N - \lp n - n_0\rp} $.

\section{Derivation of the Two Dimensional Propagator}
\label{sec:two_prop_deriv}

Solving the effective 1D Master equation
\eeqref{eq:q_master_effective}, and taking the $z$ transform gives

\begin{align}
  &\wt{\widehat{Q}}(n_{_1}, \kappa_2, z) = \frac{\lambda_1 \ls
    \e^{-\mathrm{i} \kappa_2 n_{0_2}}f_2^{-\frac{n_{0_2}}{2}} \lb 1 - f_2^{-1}\rb \rs}
{1 - z \omega \lb 1 - \frac{q_1}{2} + \frac{q_2}{2
\eta_1} \cos\ls \kappa_2 \rs \rb } \nonumber \\
&+ \frac{f_1^{\frac{n_{_1}- n_{0_1} - 1}{2}} \e^{-\mathrm{i} \kappa_2 n_{0_2}}f_2^{-\frac{n_{0_2}}{2}} \lb 1 - f_2^{-1}\rb }{N_1} \sum_{k_1 =
1}^{N_{_1}- 1} \lc f_1^{\frac{1}{2}} \sin \ls \frac{n_{_1}k_1 \pi}{N_1} \rs
- \sin \ls \lb n_{_1}- 1\rb \frac{k_1 \pi }{N_1}\rs \rc \nonumber \\
&\times \lc f_1^{\frac{1}{2}} \sin \ls \frac{n_{0_1} k_1 \pi}{N_1} \rs -
\sin \ls \lb n_{0_1} - 1\rb \frac{k_1 \pi }{N_1}\rs \rc \lc \eta_1 - 
\cos\ls \frac{k_1 \pi }{N_1}\rs \rc^{-1} \nonumber \\
&\times \lc 1 - z \lb 1 - \frac{q_1}{2} - \frac{q_2}{2} + \frac{q_1}{2
\eta_1}\cos \ls \frac{k_1 \pi_1 }{N_1}\rs + \frac{q_2}{2\eta_2} \cos\ls
\kappa_2 \rs \rb \rc ^{-1}, 
\label{eq:effective_1d_sol}
\end{align}
where
\begin{align*}
  f_2 = \frac{1 - g_2}{1+g_2}, \quad
  \eta_2 = \frac{1 + f_2}{2f_2^{\frac{1}{2}}} \quad
  \text{and} \quad
  \lambda_i = \frac{f_i^{n_i - 1} \lb 1 - f_i \rb}{1 - f_i^{N_i}}.
\end{align*}

To find the analytic expression for the 2D random walker with bias and
reflective boundaries in the first dimension, while diffusive and
unbounded in the second dimension, one needs to inverse Fourier
transform the second dimension to obtain
\label{sec:2d_prop_deriv}
\begin{flalign}
  &\wt{Q}(n_{_1}, n_{_2} z) =
    \lambda_1 \, \lc \frac{2 \eta_2 f_2^{\frac{n_{_2} - n_{0_2}}{2}} \vphi^{- \left| n_{_2} - n_{0_2} \right|}}
    {z\omega q_2 \sinh\ls \acosh\lb \frac{1}{\zeta} \rb \rs} \rc\nonumber \\
    &+ \frac{f_1^{\frac{n_{_1} - n_{0_1} - 1}{2}}}{N_1} 
    \sum_{k_1 = 1}^{N_1- 1}
    \frac{\lc f_1^{\frac{1}{2}} \sin \ls \frac{n_{_1}k_1 \pi}{N_1} \rs
- \sin \ls \lb n_{_1}- 1\rb \frac{k_1 \pi }{N_1}\rs \rc 
      \lc f_1^{\frac{1}{2}} \sin \ls \frac{n_{0_1} k_1\pi}{N_1} \rs -
\sin \ls \lb n_{0_1} - 1\rb \frac{k_1 \pi }{N_1}\rs \rc} {\eta_1
- \cos\ls \frac{k_1 \pi }{N_1}\rs } \nonumber \\
  &\times
    \frac{2 \eta_2 f_2^{\frac{n_2 - n_{0_2}}{2}}\breve{\vphi}^{- \left|  n_{_2} - n_{0_2} \right|}}
    {z\omega q_2 \sinh\ls \acosh\lb \frac{1}{\breve{\zeta}} \rb \rs},
\label{eq:bound_unbound_2d}
\end{flalign}

where (redefining)
\begin{align*}
&\zeta = \frac{z \omega q_2}{2 \eta_2 \ls 1 - z \omega \lb 1- \frac{q_2}{2} \rb \rs}, \quad
  \breve{\zeta} = \frac{z \omega q_2}{2 \eta_2 \lc 1 - z \omega \ls 1 - \frac{q_1}{2} - \frac{q_2}{2} + \frac{q_1}
  {2 \eta_1}\cos \lb \frac{k \pi_1 }{N_1}\rb \rs \rc}, \quad 
 \vphi = \exp{\ls \acosh{\lb \frac{1}{\zeta} \rb}\rs}
\end{align*}
and
\begin{align*}
\breve{\vphi} = \exp{\ls \acosh{\lb \frac{1}{\breve{\zeta}} \rb}\rs}.
\end{align*}
On \eeqref{eq:bound_unbound_2d}, we use the method of images to impose
the boundary condition following the procedure outlined in
\apref{sec:symmetrisation_two_bound} and asymmetrise the second
dimension to yield

\begin{align}
  &\wt{P}^{(r_1, r_2)}_{\vec{n}_0}(n_{_1}, n_{_2} z) =
    \frac{2\lambda_1\eta_2 f_2^{\frac{n_2 - n_{0_2}}{2}}}
    {z q_2 \sinh\ls \acosh \lb \frac{\eta_2}{\beta} \rb \rs}
    \lc \frac{\alpha^{N_2 - \lp n_2 - n_{0_2} \rp} + \alpha^{-N_2 + \lp n_2 - n_{0_2} \rp} -
    \alpha^{N_2 - \lp n_2 + n_{0_2} \rp} \xi  -\alpha^{-N_2 + \lp n_2 + n_{0_2} \rp} \xi^{-1}}
    {\sinh\ls N_2 \acosh{\lb \frac{\eta_2}{\beta} \rb} \rs} \rc, 
    \nonumber \\
&+ \frac{f_1^{\frac{n_{_1}- n_{0_1} - 1}{2}}}{N_1} \sum_{k_1 =
1}^{N_{_1}- 1} \frac{\lc f_1^{\frac{1}{2}} \sin \ls \frac{n_{_1}k_1 \pi}{N_1} \rs
- \sin \ls \lb n_{_1}- 1\rb \frac{k_1 \pi }{N_1}\rs \rc
\lc f_1^{\frac{1}{2}} \sin \ls \frac{n_{0_1} k_1 \pi}{N_1} \rs -
\sin \ls \lb n_{0_1} - 1\rb \frac{k_1\pi }{N_1}\rs \rc}{ \eta_1 -
\cos\ls \frac{k_1 \pi }{N_1}\rs} \nonumber \\
  &\times
    \frac{2\eta_2 f_2^{\frac{n_2 - n_{0_2}}{2}}}
    {z  q_2 \sinh\ls \acosh \lb \frac{\eta_2}{\breve{\beta}} \rb \rs}
\lc \frac{\breve{\alpha}^{N_2 - \lp n_2 - n_{0_2} \rp} + \breve{\alpha}^{-N_2 + \lp n_2 - n_{0_2} \rp} -
\breve{\alpha}^{N_2 - \lp n_2 + n_{0_2} \rp} \breve{\xi}  -\breve{\alpha}^{-N_2 + \lp n_2 + n_{0_2} \rp} \breve{\xi}^{-1}}
{\sinh\ls N_2 \acosh{\lb \frac{\eta_2}{\breve{\beta}} \rb} \rs} \rc, 
\end{align}

where (redefining)
\begin{flalign*}
&\beta = \frac{z q_2}{2 \ls 1 - z \lb 1- \frac{q_2}{2} \rb \rs}, \quad
\breve{\beta} = \frac{z q_2}{2 \lc 1 - z \ls 1 - \frac{q_1}{2} - \frac{q_2}{2}
+ \frac{q_1} {2 \eta_1}\cos \lb \frac{k \pi_1 }{N_1}\rb \rs \rc}, \quad 
 \alpha = \exp{\ls \acosh{\lb \frac{\eta_2}{\beta} \rb}\rs}, \\
&\breve{\alpha} = \exp{\ls \acosh{\lb \frac{\eta_2}{\breve{\beta}} \rb}\rs} \quad
\xi = \frac{f_2^{\frac{1}{2}} - \alpha}{f_2^{\frac{1}{2}} - \frac{1}{\alpha}}, \quad
\text{and} \quad
\breve{\xi} = \frac{f_2^{\frac{1}{2}} - \breve{\alpha}}{f_2^{\frac{1}{2}} - \frac{1}{\breve{\alpha}}}.
\end{flalign*}
Employing the general identity \eeqref{eq:identity_ref} (below) before
inverse $z$ transforming results in the time-dependent 2D propagator
\eeqref{eq:prop_2d_sum_t_r_1N} found in the main text.

\section{Constructing Propagators of Higher Dimensions}
\label{sec:higher_dims_construct_ap}
To build a $d$-dimensional confined lattice random walk with bias, one
first considers a semi-confined LRW where the first $d-1$ dimensions
are bounded while the final $d^{\text{th}}$ dimension is
unbounded. Symmetrising the dynamics in the $d^{\text{th}}$ dimension,
yields a biased confined LRW in the $d-1$ dimension while being
diffusive and unconfined in the $d^{\text{th}}$ dimension. By Fourier
transforming along the $d^{\text{th}}$ dimension reduces the problem
to an effective $d-1$ dimensional biased LRW whose solution is
known. Solving for the dynamics in the (diffusive) $d^{\text{th}}$
dimension in the Fourier-$z$-domain and imposing boundary condition
via the method of images, gives the solution to a confined random walk
with bias in $d-1$ dimensions and no bias in the $d^{th}$
dimension. Inverting the symmetrisation procedure along the
$d^{\mathrm{th}}$ dimension yields the confined BLRW in $d$-dimensions
in $z$-domain. Finally, with the use of the identities
\eqref{eq:identity_ref}, \eqref{eq:identity_ab},
\eeqref{eq:identity_mi} or \eqref{eq:identity_p} one inverts the
propagator from the $z$-domain to the time domain. With such a
procedure one can build propagators with arbitrary dimensions and
arbitrary boundary conditions.

\section{Identities of Finite Trigonometric Series}
\label{sec:trig_identities}
For the derivation of the higher dimensional propagators analytic identities can be
obtained by equating the $z$-transform of \eeqref{eq:1d_all_prop} for
each of the different boundary cases with the corresponding equations
\eqref{eq:p_bounded_a_1N}, \eqref{eq:p_bounded_r_1N} and \eqref{eq:p_bounded_p}. For the reflecting condition
we find
\begin{align}
  &\frac{\lc f^{\frac{1}{2}} U_{M-1 - m_>}{\ls \frac{\eta}{\gamma} \lb \gamma - \mu \rb \rs}
    - U_{M - m_>}{\ls \frac{\eta}{\gamma} \lb  \gamma - \mu \rb \rs}\rc
    \lc f^{\frac{1}{2}} U_{m_< - 1}{\ls \frac{\eta}{\gamma} \lb  \gamma - \mu \rb \rs}
    - U_{m_< - 2}{\ls \frac{\eta}{\gamma} \lb  \gamma - \mu \rb \rs}\rc}
    { \mu  U_{M-1}{\ls  \frac{\eta}{\gamma} \lb  \gamma - \mu \rb \rs}} \nonumber \\
&\equiv \frac{f^{\frac{m_1 + m_2 - 1}{2}} \lb f- 1 \rb}{\mu \lb 1 - f^M \rb } + 
\frac{1}{M}\sum_{k = 1}^{M-1}
\frac{ \lb f^{\frac{1}{2}} \sin \ls \frac{m_1 k \pi}{N} \rs - \sin \ls \lb m_1 - 1\rb
\frac{k \pi }{M}\rs \rb \lb f^{\frac{1}{2}} \sin \ls \frac{m_2 k \pi}{N} \rs - \sin \ls \lb m_2 - 1\rb
\frac{k \pi }{M}\rs \rb}{\lb \eta  - \cos\ls \frac{k \pi}{N} \rs \rb
\ls  \gamma - \mu   - \frac{\gamma}{\eta}\cos{\ls\frac{k \pi}{M}\rs}\rs},
\label{eq:identity_ref}
\end{align}
for the absorbing case we generate
\begin{align}
\frac{
U_{M-1 - m_>}{\lb \frac{\eta}{\gamma} \rb}
  U_{m_< - 2}{\lb \frac{\eta}{\gamma} \rb}}
  {U_{M - 2}{\lb \frac{\eta}{\gamma} \rb}}
\equiv \frac{1}{M-1}  \sum_{k = 1}^{M-1} \frac{
 \sin\ls \lb\frac{m_1-1}{M-1}\rb k\pi\rs \sin\ls
  \left(\frac{m_2-1}{M-1}\right)k\pi \rs}
  {\frac{\eta}{\gamma} - \cos{\ls \frac{k \pi}{M - 1} \rs}},
\label{eq:identity_ab}
\end{align}
and for the periodic domain we obtain
\begin{align}
  \frac{f^{\frac{m}{2}} \ls
    U_{M - 1 - \lp m \rp}{\lb \frac{\eta}{\gamma} \rb}
  + U_{\lp m \rp - 1}{\lb \frac{\eta}{\gamma} \rb} f^{-\frac{M \, \mathrm{sgn}\lb m \rb}{2}} \rs
    }{T_{M}{\lb \frac{\eta}{\gamma} \rb}
-T_{M}{\lb \eta \rb} } 
  \equiv \frac{1}{M} \sum_{k = 0}^{M-1}\frac{\exp \lb \frac{2 k \pi \mathrm{i}m}{M} \rb }
    {\frac{\eta}{\gamma} - \cosh{\ls \frac{2k\pi\mathrm{i}}{M} - \frac{1}{2} \ln{\lb f \rb}\rs}}.
\label{eq:identity_p}
\end{align}
In equations \eqref{eq:identity_ref}, \eqref{eq:identity_ab} and
\eqref{eq:identity_p} the meaning of the symbols are as follows:
$\gamma$ and $\mu$ are complex constants; $M, m$ and $n$ are integers
with $1 \leq m_1, m_2 \leq N$; $f > 0$;
$m_> = \frac{1}{2} \lb m_1 + m_2 + \lp m_1 - m_2\rp\rb$ and
$m_< = \frac{1}{2} \lb m_1 + m_2 - \lp m_1 - m_2\rp\rb$; and
$\eta = \frac{1}{2} (1 + f)f^{-\frac{1}{2}}$. The validity of
equations \eqref{eq:identity_ref} and \eqref{eq:identity_ab} is based
on the known general identity (E1) in reference \cite{giuggioli2020},
while \eeqref{eq:identity_p} is a new identity that reduces to (E3) in
reference \cite{giuggioli2020} when $f, \eta = 1$. There is also a
relation (numerical identity) that can be obtained from the mixed
scenario using the procedure in \apref{sec:time_dep_mixed} given by
\begin{align}
  \frac{U_{M  - 1 - m_>}{\lb \frac{\eta}{\gamma}\rb}
    \ls f^{\frac{1}{2}} U_{m_< - 1}{\lb \frac{\eta}{\gamma} \rb}-
    U_{m_< - 2}{\lb \frac{\eta}{\gamma} \rb}\rs}
  {\ls f^{\frac{1}{2}}
U_{M - 1}{\lb \frac{\eta}{\gamma} \rb}
-U_{M - 2}{\lb \frac{\eta}{\gamma} \rb} \rs } 
  \equiv \sum_{k = 1}^{M-1}
  \frac{\sin \ls \lb M - m_> \rb \theta_k \rs
  \lc  f^{\frac{1}{2}} \sin \ls   m_< \theta_k\rs -
  \sin \ls  \lb m_<  - 1\rb \theta_k \rs \rc}
  {\lc \lb M - 1 \rb \cos\ls \lb M-1\rb \theta_k\rs -
  M f^{\frac{1}{2}} \cos{\ls M \theta_k \rs}
 \rc
\ls \frac{\eta}{\gamma} - \cos{\lb \theta_k \rb} \rs
  },
\label{eq:identity_mi}
\end{align}
where $\cos{\lb \theta_k \rb}$ is the $k^{th}$ (numerical) root of the
orthogonal polynomial
$f^{\frac{1}{2}}U_{M-1}\ls \cos{\lb \theta \rb}\rs -
U_{M-2}\ls\cos{\lb \theta \rb}\rs$.

\section{Mean First-Return Times in Higher Dimensions}
\label{sec:mean_return_ap}
A hint of the non-trivial dependence of the return dynamics can be
evinced by studying how different initial positions affect the MRT. We
display for this purpose in figure~\ref{fig:return_mean_2d} the
reciprocal of the MRT in a 2D domain with reflecting boundaries with
different starting locations, $\vec{n}$ , as a function of the bias
$\vec{g}$.  Each panel from (a)-(d) represents a different starting
location which is progressively closer to the top-left corner. In the
presence of a bias, the walker is pushed away from the starting
site. For an initial location at the centre of the domain it results
in a weak dependence on the bias direction as shown in panel (a). A
strong dependence when the starting location is off-centre is instead
shown in panels (b)-(d). With the shift in the starting sites from
panel (b) to (c) to (d), there is a shorter MRT the stronger the bias
is directed towards that corner ($g_1 < 0, g_2 > 0$) with instead long
MRT for all other bias directions. In panel (b), one may also notice
an asymmetry with respect to the diagonal which is not present in
panels (c) and (d). It is due to the starting site being closer to the
top boundary at $n_2 = N_2 $ than the left boundary at $n_1 = 1$.
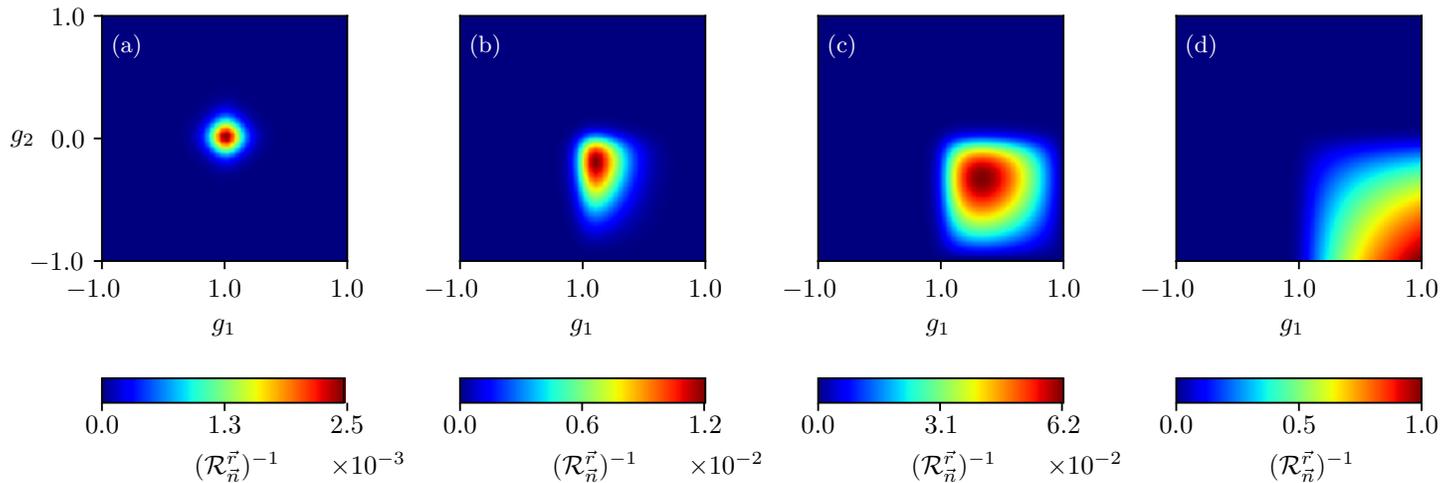
\begin{figure*}[hbt]
  \hfill
  \fontsize{10}{10}
  \begin{overpic}[tics=2, percent,width=0.95\textwidth]{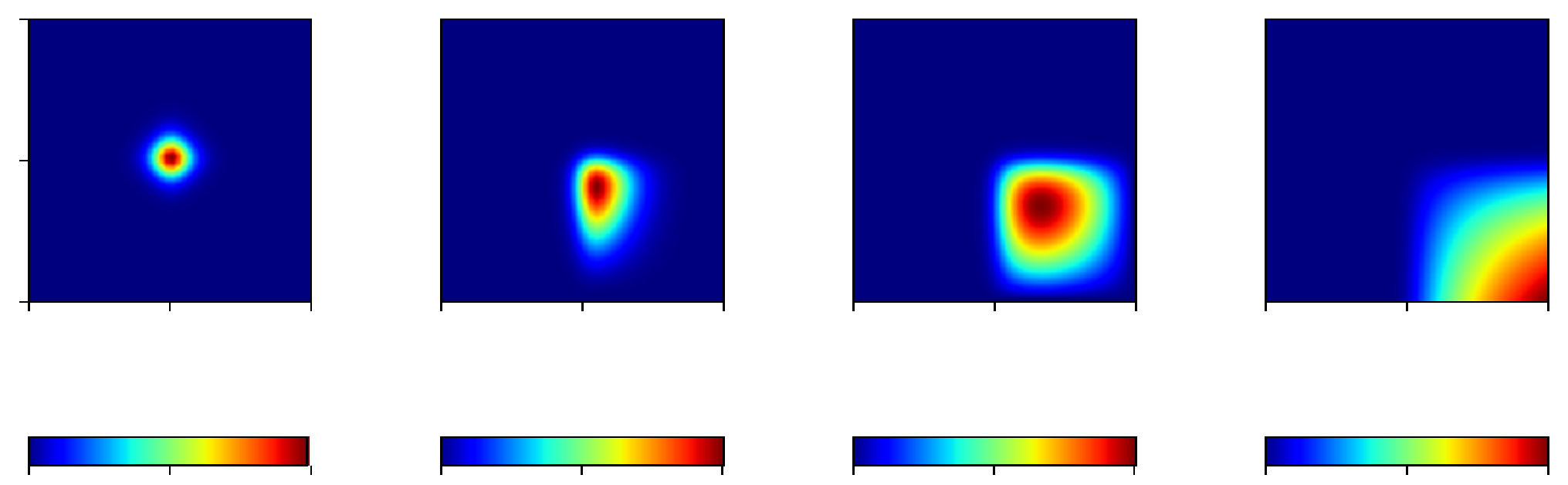}
    \put(-0.8  , 9.0){$-1.0$}
    \put(9.6   , 9.0){$1.0$}
    \put(18.6  , 9.0){$1.0$}
    \put(9.9   , 6.5){$g_1$}

    \put(25.5  , 9.0){$-1.0$}
    \put(35.9  , 9.0){$1.0$}
    \put(44.9  , 9.0){$1.0$}
    \put(36.2  , 6.5){$g_1$}

    \put(51.8  , 9.0){$-1.0$}
    \put(62.25 , 9.0){$1.0$}
    \put(71.25 , 9.0){$1.0$}
    \put(62.35  , 6.5){$g_1$}

    \put(78.1  , 9.0){$-1.0$}
    \put(88.55 , 9.0){$1.0$}
    \put(97.55 , 9.0){$1.0$}
    \put(88.35  , 6.5){$g_1$}

    \put(-3.5  , 11.0){$-1.0$}
    \put(-1.9  , 20.0){$0.0$}
    \put(-1.9  , 29.0){$1.0$}
    \put(-4.9  , 20.4){$g_2$}

    \put(2.5, 27){\color{white}(a)}
    \put(28.8, 27){\color{white}(b)}
    \put(55.1, 27){\color{white}(c)}
    \put(81.4, 27){\color{white}(d)}

    \put(0.60  , -1){$0.0$}
    \put(9.6   , -1){$1.3$}
    \put(18.6  , -1){$2.5$}
    \put(18.6  , -4){$\times 10^{-3}$}
    \put(8.6   , -4){$(\mathcal{R}_{\vec{n}}^{\vec{r}})^{-1}$}

    \put(26.9  , -1){$0.0$}
    \put(35.9  , -1){$0.6$}
    \put(44.9  , -1){$1.2$}
    \put(44.9  , -4){$\times 10^{-2}$}
    \put(34.9  , -4){$(\mathcal{R}_{\vec{n}}^{\vec{r}})^{-1}$}

    \put(53.2  , -1){$0.0$}
    \put(62.25 , -1){$3.1$}
    \put(71.25 , -1){$6.2$}
    \put(71.25 , -4){$\times 10^{-2}$}
    \put(61.25 , -4){$(\mathcal{R}_{\vec{n}}^{\vec{r}})^{-1}$}

    \put(79.5  , -1){$0.0$}
    \put(88.55 , -1){$0.5$}
    \put(97.55 , -1){$1.0$}
    \put(87.55 , -4){$(\mathcal{R}_{\vec{n}}^{\vec{r}})^{-1}$}
  \end{overpic}
  \vspace{30pt}
  \caption{(Colour Online) The reciprocal of the MRT,
    $\lb \mathcal{R}^{(\vec{r})}_{\vec{n}} \rb^{-1}$, of a 2D BLRW
    with domain size $\vec{N} = (20, 20)$ and diffusion parameter
    $\vec{q} = (0.8, 0.8)$ as a function of the bias $\vec{g}$. Each
    panel from (a) to (d) represents different starting sites, respectively,
    $\vec{n} = (10, 10), (5, 18), (2, 19)$ and $(1, 20)$. A positive
    (negative) $g_1$ indicates a drift to the left (right), while a
    positive (negative) $g_2$ indicates a drift downwards (upwards).}
\label{fig:return_mean_2d} 
\end{figure*}

All panels in figure~\ref{fig:return_mean_2d} display dependence on
the bias strength. When the starting location is near the centre, the
bias towards a corner yields long MRTs when compared with a diffusive
walker ($g_1 = g_2 = 0$) which has a natural tendency to stay near the
starting location. Conversely, with the starting location at a corner,
panel (d), one finds the shortest MRT when the walker is kept at the
starting location with the bias $\vec{g} = (1, -1)$. Interestingly,
when the starting location is off-centre and not at the boundary
corner (panels (b) and (d)) the MRT is minimised for an intermediate
bias strength. The latter is strong enough to reduce the number of
trajectories travelling right or downwards from the starting site
whilst weak enough to allow the walker to travel against the bias when
near the top-left corner. The precise location of the minimum can be
computed numerically for arbitrary dimensions from the explicit
definition
$\mathcal{R}_{\vec{n}}^{(\vec{r})} = \ls \prod_{j = 1}^{(d)}
h_{0_j}^{(\vec{r})}(n_{_j},n_{_j}) \rs^{-1}$.
\end{widetext}
\end{document}